\documentclass[11pt]{article}
\usepackage{amssymb}
\usepackage{amsfonts}
\usepackage{graphicx}
\usepackage{amsmath}

\makeatletter \@addtoreset{equation}{section} \makeatother

\newtheorem{theorem}{Theorem}
\newtheorem{lemma}{Lemma}
\newtheorem{corollary}{Corollary}
\newtheorem{remark}{Remark}

\newtheorem{proposition}{Proposition}

\def\J{{\cal J}}

\def\e{\varepsilon}

\def\ti{\tilde}
\def\intd{\displaystyle\int}
\def\fracd{\displaystyle\frac}
\def\sumd{\displaystyle\sum}

\begin{document}

\title{On  Universality  for Orthogonal Ensembles of Random Matrices}
\author{ M. Shcherbina\\
 Institute for Low Temperature Physics, Kharkov,
Ukraine. \\E-mail: shcherbi@ilt.kharkov.ua
}

\date{}

\maketitle

\begin{abstract}
We prove universality of local eigenvalue statistics in the bulk of the
spectrum for orthogonal invariant matrix models
with real analytic potentials with  one interval  limiting spectrum.
Our starting point is  the Tracy-Widom formula for the matrix reproducing kernel.
The key idea of the proof is to represent  the differentiation operator matrix
written  in the basis of orthogonal polynomials
as a product of a positive Toeplitz matrix and a two diagonal skew symmetric
Toeplitz matrix.

\end{abstract}

\section{Introduction and main result}\label{sec:1}

In this paper we consider ensembles of $n\times n$ real symmetric
(or Hermitian ) matrices $M$  with the probability distribution
\begin{equation}
P_{n}(M)dM=Z_{n,\beta}^{-1}\exp \{-\frac{n\beta}{2}\mathrm{Tr}V(M)\}dM,  \label{p(M)}
\end{equation}
where $Z_{n,\beta}$ is the normalization  constant, $V:\mathbb{R}\to \mathbb{R%
}_{+}$ is a H\"{o}lder function satisfying the condition
\begin{equation}\label{condV}
|V(\lambda )|\geq 2(1+\epsilon )\log(1+ |\lambda |).
\end{equation}
A positive parameter $\beta$ here assumes the values $\beta=1$ (in
the case of real symmetric matrices) or $\beta=2$ (in the Hermitian
case), and $dM$ means the  Lebesgue measure on the algebraically
independent entries of $M$. Ensembles of random matrices
(\ref{p(M)}) in the real symmetric case are usually called
orthogonal, and in the Hermitian case - unitary ensembles. This
terminology reflects the fact that the density of (\ref{p(M)}) is
invariant with respect to the orthogonal, or unitary transformation of
matrices $M$.

The
joint eigenvalue distribution corresponding to (\ref{p(M)}) has the form (see \cite{Me:91})
\begin{equation}
p_{n,\beta}(\lambda_1,...,\lambda_n)=Q_{n,\beta}^{-1}\prod_{i=1}^n
e^{-n\beta V(\lambda_i)/2}\prod_{1\le j<k\le
n}|\lambda_i-\lambda_j|^\beta, \label{p(l)}
\end{equation}
where $Q_{n,\beta}$ is the normalization constant. The simplest question  in both cases ($\beta=1,2$)
is the behavior of the
eigenvalue counting measure (NCM) of the matrix. According to
\cite{BPS,Jo:98} the NCM tends weakly in probability, as $n\to
\infty $, to the non random limiting measure ${\cal N}$ known as the
Integrated Density of States (IDS) of the ensemble, which is one of the main outputs
of studies of the global regime. The IDS is
normalized to unity and it is absolutely continuous, if $V'$ satisfies
the Lipshitz condition  \cite{Sa-To:97}.
The non-negative density $\rho(\lambda) $  is called  the
 Density of States (DOS) of the ensemble. The IDS can be found as a
unique solution of a certain variational problem (see
\cite{BPS,De:98,Sa-To:97}).

Local regimes, or local eigenvalue statistics for unitary ensembles are also well
studied now. The  problem
is to study the behavior of marginal densities
\begin{equation}\label{p_nl}
p^{(n)}_{l,\beta}(\lambda_1,...,\lambda_l)=
\int_{\mathbb{R}^{n-l}} p_{n,\beta}(\lambda_1,...\lambda_l,\lambda_{l+1},...,\lambda_n)
d\lambda_{l+1}...d\lambda_n
\end{equation}
in the scaling limit, when
$\lambda_i=\lambda_0+s_i/n^\kappa$ $(i=1,\dots,l)$,
and $\kappa$ is a constant, depending on the behavior of the limiting
density $\rho(\lambda)$ in a small neighborhood of $\lambda_0$ of the limiting spectrum $\sigma$.
If $\rho(\lambda_0)\not=0$, then $\kappa=1$, if $\rho(\lambda_0)=0$
and $\rho(\lambda)\sim |\lambda-\lambda|^\alpha$, then $\kappa=1/(1+\alpha)$.
The universality conjecture states that  the scaling limits of all marginal densities
 are universal, i.e. do not depend on $V$.

In the case of unitary ensembles all marginal densities can be expressed
 (see \cite{Me:91}) in terms of the unique function $K_{n,2}(\lambda,\mu)$.
\begin{equation} \label{p_k=}
p^{(n)}_{l,\beta}(\lambda_1,...,\lambda_l)=
\frac{(n-l)!}{n!}\det \{K_{n,2}(\lambda
_{j},\lambda _{k})\}_{j,k=1}^{l}.
\end{equation}%
This function has the form
\begin{equation}\label{k_n}
K_{n,2}(\lambda ,\mu )=\sum_{l=0}^{n-1}\psi _{l}^{(n)}(\lambda )\psi
_{l}^{(n)}(\mu )
\end{equation}%
and is known as a reproducing kernel of the orthonormalized system
\begin{equation}\label{psi}
\psi _{l}^{(n)}(\lambda )=\exp \{-n V(\lambda )/2\}p_{l}^{(n)}(\lambda
),\;\,l=0,...,
\end{equation}
in which $\{p_{l}^{(n)}\}_{l=0}^n$ are orthogonal polynomials on $\mathbb{R}$
associated with the weight
$ w_{n}(\lambda )=e^{-n V(\lambda )}$ i.e.,
\begin{equation}\label{ortP}
\int p_{l}^{(n)}(\lambda )p_{m}^{(n)}(\lambda )w_{n}(\lambda )d\lambda
=\delta _{l,m}.
\end{equation}%
Hence, the problem to study  marginal distributions is replaced by the problem
to study the behavior of the reproducing kernel $K_n(\lambda,\mu)$ in the scaling limit.

This problem was solved in  many cases. For example, in the
 bulk case ($\rho(\lambda_0)\not=0$) it was shown in
\cite{PS:97}, that for a  general class of $V$ (the third
derivative is bounded in the some neighborhood of $\lambda_0$)
\[
\lim_{n\to\infty}\frac{1}{n\rho(\lambda_0)}K_{n,2}(\lambda_0+s_1/n\rho(\lambda_0),
\lambda_0+s_2/n\rho(\lambda_0))=\mathcal{K}_{\infty,2}^{(0)}(s_1,s_2),\]
where $\mathcal{K}_0(s_1,s_2)$ is a universal $sin$-kernel
\begin{equation}\label{K_0}
\mathcal{K}_{\infty,2}^{(0)}(s_1-s_2)=\frac{\sin\pi(s_1-s_2)}{\pi(s_1-s_2)}.
\end{equation}
This result for the case of real analytic $V$ was obtained also in
\cite{De:99}.

 For unitary ensembles it is also possible to study (see \cite{De:99}) the edge universality,
i.e. the case when $\lambda_0$ is the edge point of the spectrum
and $\rho(\lambda)\sim|\lambda-\lambda_0|^{1/2}$, as
$\lambda\sim\lambda_0$. There are also results on the extreme
point universality (double scaling limit). This means
universality of the limiting kernel in the case when
$\rho(\lambda)\sim (\lambda-\lambda_0)^2$, as
$\lambda\sim\lambda_0$. See \cite{CK} for the result for real
analytic potentials and \cite{S:05} for a general case.

 For orthogonal ensembles ($\beta=1$ ) the situation is more complicated. Instead of
(\ref{k_n})
we have to use $2\times 2$ matrix kernel
\begin{equation}\label{K_n1}
K_{n,1}(\lambda,\mu)=\left(\begin{array}{cc}S_n(\lambda,\mu)&
S_nd(\lambda,\mu)\\ IS_n(\lambda,\mu)-\epsilon(\lambda-\mu)& S_n(\mu,\lambda)\end{array}\right).
\end{equation}
Here $S_n(\lambda,\mu)$ is some scalar kernel (see (\ref{S}))
below), $d$ denotes the differentiating, $IS_n(\lambda,\mu)$
can be obtained from $S_n$ by some integration procedure and $\epsilon(\lambda)$
is defined in (\ref{eps}).
Similarly to the unitary case all marginal densities can be
expressed in terms of the kernel $K_{n1}$ (see \cite{Tr-Wi:98}),
e.g.
\[\rho_n(\lambda)=p^{(n)}_{1,1}(\lambda)=\frac{1}{2n}\hbox{Tr}K_{n,1}(\lambda,\lambda),
\]
and
\[p_{2,1}^{(n)}(\lambda,\mu)=\frac{1}{4n(n-1)}\left[\hbox{Tr}
K_{n,1}(\lambda,\lambda)
\hbox{Tr}K_{n,1}(\mu,\mu)-2\hbox{Tr}\left(K_{n,1}(\lambda,\mu)K_{n,1}(\mu,\lambda)\right)\right].
\]
The matrix kernel  (\ref{K_n1}) was introduced first in
\cite{Dy} for  circular ensemble and then in \cite{Me:91}  for
 orthogonal ensembles. The scalar kernels of (\ref{K_n1}) could
be defined in principle in terms of any family  of polynomials complete in
$L_2(\mathbb{R},w_n)$  (see
\cite{Tr-Wi:98}), but usually the families of skew orthogonal
polynomials were used (see \cite{Me:91} and references therein).
Unfortunately, for general weights the properties of skew
orthogonal polynomials are not studied enough. Hence,  using of
skew orthogonal polynomials for $V$ of a general type rises
serious technical difficulties.
In the paper \cite{DeG1} a new approach to this problem was
proposed. It is based on the result of \cite{Tr-Wi:98}, which
allows to express the kernel $S_n(\lambda,\mu)$ in terms of the
family of orthogonal polynomials (\ref{ortP}). Using the
representation of \cite{Tr-Wi:98}, it was shown that
  $S_n(\lambda,\mu)\to K_{\infty,2}^{(0)}(\lambda,\mu)$,
where $K_{\infty,2}^{(0)}(\lambda,\mu)$ is defined by (\ref{K_0}).
The same approach was used in \cite{DeG2} to prove the edge universality.
Unfortunately,   the papers \cite{DeG1} and \cite{DeG2} deal only with the case,
when (in our notations) $V(\lambda)=\lambda^{2m}+
n^{-1/2m}a_{2m-2}\lambda^{2m-2}+\dots$.
 But since, like usually (see \cite{PS:97}, \cite{PS:07}), the small terms
 $n^{-1/2m} a_{2m-2}\lambda^{2m-2}+\dots$ have no influence on the limiting
behavior of $K_n(\lambda,\mu)$, this result in fact proves
universality for the case of monomial $V(\lambda)=\lambda^{2m}$.
 In the papers \cite{St1,St2} the bulk and the edges universality  were studied
for the case of $V$ being an even quatric polynomial.

In the present paper we prove universality in the bulk of the spectrum
for any real analytic $V$ with one interval support.

Let us state  our main conditions.

\begin{description}

\item[\textbf{ C1.} ] \textit{$V(\lambda )$ satisfies (\ref{condV}) and
is an even analytic
function in }
\begin{equation}
\Omega[d_1,d_2]=\{z:-2-d_1\le\Re z\le 2+d_1,\,\, |\Im z|\le d_2\}, \quad
d_1,d_2>0.
\label{Omega}\end{equation}
 \item[\textbf{ C2.}]\textit{  The support $\sigma $ of
IDS of the ensemble consists of a single interval:}
\[
\sigma =[-2,2].
\]
\item[\textbf{C3.}] \textit{ DOS $\rho(\lambda)$ is
strictly positive in the internal points $\lambda\in (-2,2)$ and $\rho(\lambda)\sim
|\lambda\mp 2|^{1/2}$, as $\lambda\sim\pm2$}.
\item[\textbf{C4.}]\textit{ The function
\begin{equation}
u(\lambda)=2\int\log |\mu-\lambda|\rho(\mu)d\mu-V(\lambda)
\label{u}\end{equation}
 achieves its maximum
if and only if $\lambda\in\sigma$. }
\end{description}
It is proved in \cite{APS:01}, that under conditions $C1- C4$, if
 we consider a semi infinite Jacoby matrix $\mathcal{J}^{(n)}$,
generated by the recursion relations for
 the system of orthogonal polynomials (\ref{ortP})
\begin{equation}\label{rec}
J_{l}^{(n)}\psi _{l+1}^{(n)}(\lambda )+q_{l}^{(n)}\psi _{l}^{(n)}(\lambda )+
J_{l-1}^{(n)}\psi _{l-1}^{(n)}(\lambda )=\lambda \psi _{l}^{(n)}(\lambda
),\quad J_{-1}^{(n)}=0,\quad l=0,...,
\end{equation}%
then,  $q^{(n)}_l=0$ and there exists some fixed $\gamma$
such that uniformly in $k:|k|\le 2n^{1/2}$
\begin{equation}\label{J_k}
  \bigg|J^{(n)}_{n+k}-1-\frac{k}{n}\gamma\bigg|\le C\frac{|k|^2+n^{2/3}}{n^{2}}.
\end{equation}
\begin{remark} \label{rem:4}
The convergence $J^{(n)}_{n+k}\to 1$ ($n\to\infty$)
 without  uniform bounds for the remainder terms was shown in \cite{APS:97} under much more
weak conditions ($V'(\lambda)$ is a H\"{o}lder function in some neighborhood of the
limiting spectrum).
\end{remark}
Note also (see \cite{APS:01}) that under conditions $C1-C4$ the limiting
density of states (DOS) $\rho $ has the form
\begin{equation}\label{rho}
\rho (\lambda )=\frac{1}{2\pi }P(\lambda
)\sqrt{4-\lambda^2}\,\mathbf{1}_{|\lambda|<2},
\end{equation}
where the function $P$ can be represented in the form
\begin{equation}
P(z)=\frac{1}{2\pi i }\oint_{\mathcal{L} }\frac{V^{\prime }(z)-V^{\prime
}(\zeta )}{(z-\zeta)(\zeta^2-4)^{1/2}}d\zeta =
\frac{1}{2\pi }\int_{-\pi}^{\pi}{\frac{V^{\prime }(z)-V^{\prime
}(2\cos y )}{z-2\cos y}}dy .  \label{P}
\end{equation}%
Here the contour $\mathcal{L}\subset \Omega[d_1/2,d_2/2]$ and
$\mathcal{L}$ contains inside the interval $(-2,2)$.
If $V$ is a polynomial of $2m$th degree, then it is evident that $P(z)$
is a polynomial of $(2m-2)$th degree, and conditions C3 and (\ref{aprV}) guarantee that
\begin{equation}
  \label{posP}
 |P(z)|\le C,\quad z\in\Omega[d_1/2,d_2/2], \quad P(\lambda)\ge\delta>0,\quad
 \lambda\in[-2,2].
\end{equation}
An important role below belongs to the following two operators:
\begin{equation}\label{calP}
 P_{j,k}=\frac{1}{2\pi}\int_{-\pi}^\pi P(2\cos y)e^{i(j-k)y}dy=
\frac{1}{2\pi i} \oint_{|\zeta|=1} P(\zeta+\zeta^{-1})\zeta ^{j-k-1}d\zeta
\end{equation}
and  $\mathcal{R}=\mathcal{P}^{-1}$ which has the entries:
\begin{equation}\label{R}
 R_{j,k}=R_{j-k}=\frac{1}{2\pi }\int_{-\pi}^{\pi}\frac{e^{i(j-k)x}dx}{P(2\cos x)}
 =\frac{1}{2\pi i}\oint\frac{\zeta^{-1}\zeta^{j-k}d\zeta}{P(\zeta+\zeta^{-1})}.
\end{equation}
It is important for us that
\begin{equation}\label{b_R}
\delta_1\le\mathcal{R}\le\delta_2,\quad \delta_1=\inf_{\sigma} P^{-1}(\lambda),\quad
\delta_2=\sup_{\sigma} P^{-1}(\lambda).
\end{equation}
Remark also, that if we denote by $\mathcal{J}^*$ an infinite Jacobi matrix with  constant coefficients
\begin{equation}\label{J^*}
 \mathcal{J}^{*}=\{J^*_{j,k}\}_{j,k=-\infty}^{\infty},\quad J^*_{j,k}=\delta_{j+1,k}
 +\delta_{j-1,k},
\end{equation}
then the spectral theorem yields that $\mathcal{P}=P(\mathcal{J}^*)$,
$\mathcal{R}=P^{-1}(\mathcal{J}^*)$.

Following the approach of \cite{Tr-Wi:98} and \cite{DeG1}, we consider
\begin{equation}\label{eps}
\epsilon(\lambda)=\frac{1}{2}\hbox{sign}(\lambda);\quad \epsilon
 f(\lambda)=\int\epsilon(\lambda-\mu)f(\mu)d\mu;
\end{equation}
\begin{equation}\label{M}
M_{j,l}=n(\psi^{(n)}_j,\epsilon\psi^{(n)}_l);\quad
\mathcal{M}^{(0,\infty)}=\{M_{j,l}\}_{j,l=0}^\infty;\quad
\mathcal{M}^{(0,n)}=\{M_{j,l}\}_{j,l=0}^{n-1}.
\end{equation}
Then, according to \cite{Tr-Wi:98}, the kernel $S_n(\lambda,\mu)$ has the
form:
\begin{equation}\label{S}
S_n(\lambda,\mu)=-\sum_{i,j=0}^{n-1}\psi^{(n)}_i(\lambda)(\mathcal{M}^{(0,n)})^{-1}_{i,j}
(n\epsilon\psi^{(n)}_j)(\mu).
\end{equation}
The main result of the paper is
\begin{theorem}\label{thm:1}
Consider the orthogonal ensemble of random matrices defined by (\ref{p(M)})-(\ref{p(l)})
with  $V$ satisfying conditions  C1-C4. Then for
$\lambda_0$ in the bulk ($\rho(\lambda_0)\not=0$) there exist weak limits
of the scaled correlation functions (\ref{p_nl}) and these limits
are given  in terms of the universal matrix kernel
\begin{equation}\label{bulk}
K^{(0)}_{\infty,1}(s_1,s_2)=\lim_{n\to\infty}\frac{1}{n\rho(\lambda_0)}K_{n,1}(\lambda_0+s_1/n\rho(\lambda_0),
\lambda_0+s_1/n\rho(\lambda_0)),
\end{equation}
where  $K_{n,1}(\lambda,\mu)$ is defined by (\ref{K_n1})-(\ref{S}), and
\[K_{\infty,1}^{(0)}(s_1,s_2)=\left(\begin{array}{cc}
K_{\infty,2}^{(0)}(s_1-s_2)& \frac{\partial}{\partial s_1}K_{\infty,2}^{(0)}(s_1-s_2)\\
\int_0^{s_1-s_2}K_{\infty,2}^{(0)}(t)dt-\epsilon(s_1-s_2)&K_{\infty,2}^{(0)}(s_1-s_2)
\end{array}\right),
\]
with $K_{\infty,2}^{(0)}(s_1-s_2)$ of the form (\ref{K_0}).
\end{theorem}
 The proof of the theorem is based on the following result
\begin{theorem}\label{thm:2}
Under conditions of Theorem \ref{thm:1} for even $n$ the matrix $(\mathcal{M}^{(0,n)})^{-1}$
defined in (\ref{M}) is bounded uniformly in $n$, i.e. $||(\mathcal{M}^{(0,n)})^{-1}||\le C$
where $C$ is independent of $n$ and $||.||$ is a standard norm for $n\times n$ matrices.
\end{theorem}

The paper is organized as follows. In  Section \ref{sec:2} we prove Theorems 1 and 2.
The proofs of auxiliary results are given in Section \ref{sec:3}.

\section{Proof of the main results}\label{sec:2}

\noindent{\it Proof of Theorem \ref{thm:2}.} According to the
results of  \cite{APS:01} and \cite{PS:03}, if we restrict the
integration in (\ref{p_nl}) by $|\lambda_i|\le L=2+d_1/2$, consider
the polynomials $\{p^{(n,L)}_k\}_{k=0}^\infty$ orthogonal on the interval
$[-L,L]$ with the weight
$e^{-nV}$ and set $\psi^{(n,L)}_k=e^{-nV/2}p^{(n,L)}_k$,
then for $k\le n(1+\varepsilon)$ with some $\varepsilon>0$
\begin{equation}\label{apr_pol}\begin{array}{l}
\sup_{|\lambda|\le L}|\psi^{(n,L)}_k(\lambda)-
\psi^{(n)}_k(\lambda)|\le e^{-nC},\quad
|\psi^{(n)}_k(\pm L)|\le e^{-nC}
\end{array}\end{equation}
 with some
absolute $C$. Therefore  from the very beginning we can take all
integrals in (\ref{p_nl}), (\ref{ortP}), (\ref{eps}) and (\ref{M})
over the interval $[-L,L]$. Besides, observe that
since $V$ is an analytic function in $\Omega[d_1,d_2]$ (see
(\ref{Omega})), for any natural $m$ there exists a polynomial
$V_m$ of the $(2m)$th degree such that
\begin{equation}\label{aprV}
|V_m(z)|\le C_0,\quad |V(z)-V_m(z)|\le e^{-Cm},\quad z\in\Omega[d_1/2,d_2/2].
\end{equation}
Here and everywhere below we denote by $C,C_0,C_1,...$ positive $n,m$-independent constants (different
in  different formulas).

Take
\begin{equation}\label{m=}
m=[\log^2 n]
\end{equation}
and consider the system of polynomials
$\{p^{(n,L,m)}_k\}_{k=0}^\infty$  orthogonal in the interval $[-L,L]$
with respect to the weight $e^{-nV_m(\lambda)}$. Set
$\psi^{(n,L,m)}_k=p^{(n,L,m)}_ke^{-nV_m/2}$ and construct
$\mathcal{M}^{(0,n)}_m$ by (\ref{M}) with $\psi^{(n,L,m)}_k$.
Then for any $k\le n+2n^{1/2}$ and uniformly in $\lambda\in[-L,L]$
\begin{equation}\label{dif_M}\begin{array}{l}
|\psi^{(n,L)}_k(\lambda)-\psi^{(n,L,m)}_k(\lambda)|
\le e^{-C\log^2n},\quad |\varepsilon\psi^{(n,L)}_k(\lambda)-\varepsilon\psi^{(n,L,m)}_k(\lambda)|
\le e^{-C\log^2n}\\
 ||\mathcal{M}^{(0,n)}_m-\mathcal{M}^{(0,n)}||\le e^{-C\log^2n}
\end{array}\end{equation}
The proof of the first bound here is identical to the proof of (\ref{apr_pol})
(see \cite{PS:03}). The second bound follows from the first one because the operator
$\varepsilon:L_2[-L,L]\to C[-L,L]$ is bounded by $L$. The last bound in
(\ref{dif_M}) follows from the first one and the inequality valid
for the norm of an arbitrary matrix $\mathcal{A}$
\begin{equation}\label{b_norm}
||\mathcal{A}||^2\le\max_i\sum_j|A_{i,j}|\cdot\max_j\sum_i|A_{i,j}|.
\end{equation}
Remark also  that if for arbitrary matrices $\mathcal{A}$, $\mathcal{B}$
 $||\mathcal{A}^{-1}||\le C$
and $||\mathcal{A}-\mathcal{B}||\le qC^{-1}$ with some $0<q<1$, then we can write
$\mathcal{B}=
\mathcal{A}(I-\mathcal{A}^{-1}(\mathcal{A}-\mathcal{B}))$.
Since $||\mathcal{A}^{-1}(\mathcal{A}-\mathcal{B})||\le q<1$,
$||(I-\mathcal{A}^{-1}(\mathcal{A}-\mathcal{B}))^{-1}||\le(1-q)^{-1}$,
(see any textbook on linear algebra). Thus
 $\mathcal{B}$ has an inverse matrix  and $||\mathcal{B}^{-1}||\le C(1-q)^{-1}$.
Moreover, $||\mathcal{A}^{-1}-\mathcal{B}^{-1}||\le q(1-q)^{-1}C^{2}$.
Using this simple observation  and  (\ref{dif_M}), we obtain that if
$||(\mathcal{M}_m^{(0,n)})^{-1}||\le C_1$, then $||(\mathcal{M}^{(0,n)})^{-1}||\le
C_1(1- C_1e^{-C\log^2n})^{-1}\le 2C_1$ and
\[||(\mathcal{M}^{(0,n)})^{-1}-(\mathcal{M}_m^{(0,n)})^{-1}||\le C_2e^{-C\log^2n}.\]
Using this bound combined with the first and the second bound of (\ref{dif_M})
we can compare each term of the kernel $S_{n,m}(\lambda,\mu)$ constructed by
formula (\ref{S}) with  new orthogonal polynomials
$\{p^{(n,L,m)}_k\}_{k=0}^\infty$ with the corresponding
term of $S_{n}(\lambda,\mu)$. Then, since by the  result of \cite{PS:97}
\[|\psi_k^{(n)}(\lambda)|^2\le K_{n,2}(\lambda,\lambda)\le nC,\quad \lambda\in[-L,L],\]
and by the Schwarz inequality
\[|\epsilon\psi_k^{(n)}(\lambda)|\le (2L)^{1/2}||\psi_k^{(n)}||_2
\le(2L)^{1/2},\quad \lambda\in[-L,L],\]
where $||.||_2$ is a standard norm in $L_2[-L,L]$, we obtain that uniformly in
$\lambda,\mu\in[-L,L]$
\begin{equation}\label{aprM}\begin{array}{l}
|S_{n,m}(\lambda,\mu)-S_{n}(\lambda,\mu)|\le Cn^4 e^{-C\log^2n}\le e^{-C'\log^2n} .
\end{array}\end{equation}
Therefore below we will study $\mathcal{M}^{(0,n)}_m$ and $S_{n,m}(\lambda,\mu)$ instead of
$\mathcal{M}^{(0,n)}$ and $S_{n}(\lambda,\mu)$. To simplify notations we omit the indexes
$m,L$, but keep the dependence on $m$ in the estimates.

\smallskip

 Let us set our main notations. We
denote by $\mathcal{H}=l_2(-\infty,\infty)$ a Hilbert space of all
infinite sequences $\{x_i\}_{i=-\infty}^\infty$ with a standard
scalar product $(.,.)$ and a norm $||.||$. Let also $\{e_i\}_{i=-\infty}^\infty$ be
a standard basis in $\mathcal{H}$ and $\mathcal{I}^{(n_1,n_2)}$  with $-\infty\le n_1<n_2\le\infty$
be an orthogonal projection operator defined as
\begin{equation}\label{E}
\mathcal{I}^{(n_1,n_2)}e_i=\left\{\begin{array}{ll}e_i,&n_1\le i<n_2,\\
0,& otherwise.\end{array}\right.
\end{equation}
For any infinite or semi infinite matrix $\mathcal{ A}=\{A_{i,j}\}$
we will denote by
\begin{equation}\label{A_k,N}\begin{array}{l}
\mathcal{ A}^{(n_1,n_2)}=\mathcal{I}^{(n_1,n_2)}\mathcal{A }\mathcal{I}^{(n_1,n_2)},\\
(\mathcal{ A}^{(n_1,n_2)})^{-1}=
\mathcal{I}^{(n_1,n_2)}\bigg(I-\mathcal{I}^{(n_1,n_2)}+\mathcal{ A}^{(n_1,n_2)}\bigg)^{-1}
\mathcal{I}^{(n_1,n_2)},
\end{array}\end{equation}
so that $(\mathcal{A }^{(n_1,n_2)})^{-1}$ is a block operator
which is inverse to $\mathcal{ A}^{(n_1,n_2)}$ in the space
$I^{(n_1,n_2)}\mathcal{H}$ and zero on the
$(I-I^{(n_1,n_2)})\mathcal{H}$.
We denote also by $(.,.)_2$ and $||.||_2$ a standard scalar product
and a norm in $L_2[-L,L]$.

Set $\mathcal{V}^{(0,\infty)}=\{\mathcal{V}_{j,l}\}_{j,l=0}^\infty,$
where
\begin{equation}\label{nu}\mathcal{V}_{j,l}=\hbox{ sign}(l-j)(\psi^{(n)}_j,V'\psi^{(n)}_l)_2=
\frac{2}{n}\left\{\begin{array}{ll}
(\psi^{(n)}_j,
(\psi^{(n)}_l)')_2,&j>l,\\
(\psi^{(n)}_j,
(\psi^{(n)}_l)')_2+O(e^{-C\log^2n}),&j\le l.
\end{array}\right.\end{equation}
Here $O(e^{-C\log^2n})$ appears because of the integration by parts and bounds (\ref{apr_pol}),
(\ref{dif_M}).
Since $(\psi^{(n)}_k)'=q_{k}e^{-nV/2}$, where $q_{k}$ is a polynomial
of the $(k+2m-1)$th degree, its Fourier expansion in the basis $\{\psi^{(n)}_k\}_{k=0}^{\infty}$
contains not more than
$(k+2m-1)$ terms and for $|j-k|>2m-1$ the $j$th coefficient is $O(e^{-C\log^2n})$.
Therefore for $k\le n+2n^{1/2}$
\begin{equation}\label{psi'}
n^{-1}(\psi^{(n)}_k)'=\frac{1}{2}\sum_{j}\mathcal{V}_{j,k}\psi^{(n)}_j+O_2(e^{-C\log^2n}).
\end{equation}
Here and below we write $\phi(\lambda)=O_2(\varepsilon_n)$, if
$||\phi||_2\le C\varepsilon_n$. The above relation implies
 \begin{equation}\label{nu*psi}
\frac{1}{2}\epsilon\bigg(\sum_j\mathcal{V}^{(0,\infty)}_{j,k} \psi^{(n)}_j\bigg)=n^{-1}\psi^{(n)}_k
+O_2(e^{-C\log^2n}).
\end{equation}
Hence, by (\ref{M}), for $0\le j,k\le n+2n^{1/2}$
\begin{equation}\label{M*V}
\frac{1}{2}(\mathcal{M}^{(0,\infty)}\mathcal{V}^{(0,\infty)})_{j,k} =\delta_{j,k}
+O(e^{-C\log^2n}).
\end{equation}
Thus,
\begin{equation}\label{M_n*V}
 \frac{1}{2}\mathcal{M}^{(0,n)}\mathcal{V}^{(0,n)}=I^{(0,n)}-
  \mu^{(0,n)}\nu^{(0,n)}
 +\mathcal{E}^{(0,n)},\quad ||\mathcal{E}^{(0,n)}||=O(e^{-C\log^2n}),
\end{equation}
where $\nu^{(0,n)}$  is a matrix with entries equal to zero except
the block $(2m-1)\times(2m-1)$ in the right bottom corner. The block has
 the form
 \begin{equation}\label{V^n}
\nu^{(m)}=\left( \begin{array}{ccccc}
\mathcal{V}_{n,n-2m+1}&0&\mathcal{V}_{n,n-2m+3}&\dots&\mathcal{V}_{n,n-1}\\
0&\mathcal{V}_{n+1,n-2m+2}&0&\dots&0\\
0&0&\mathcal{V}_{n+2,n-2m+3}&\dots&\mathcal{V}_{n+2,n-1}\\
\dots&\dots&\dots&\dots&\dots\\
0&0&0&\dots&\mathcal{V}_{n+2m-2,n-1}.
\end{array}\right)\end{equation}
$\mu^{(0,n)}$ in (\ref{M_n*V}) has $(n-2m+1)$ first columns equal to zero and
 the last $(2m-1)$ ones of the form
\[
\mu^{(0,n)}_{l,n-2m-1+k}=M_{l,n-1+k}, \quad k=1,\dots,2m-1,\quad l=0,\dots, n-1.
\]
The relation (\ref{M_n*V}) was obtained first in \cite{DeG1}. Applying (\ref{M_n*V})
to any vector $x\in\mathcal{I}^{(0,n)}\mathcal{H}$,
we have
\begin{equation}\label{M_n*Vx}
\frac{1}{2}\mathcal{M}^{(0,n)}\mathcal{V}^{(0,n)}x=x-\sum_{k=1}^{2m-1}(x,e_{n-k})
f_k+\mathcal{E}^{(0,n)}x,\end{equation}
where
\[f_k=(f_{k,0},\dots,f_{k,n-1}),
\quad f_{k,j}=(\mu^{(0,n)}\nu^{(0,n)})_{n-k,j},\quad
||\mathcal{E}^{(0,n)}x||\le e^{-C\log^2n}||x||\]

If we make the operation of transposition of matrices  in
(\ref{M_n*V}) and apply the result to any
$x\in\mathcal{I}^{(0,n)}\mathcal{H}$, we get
\begin{equation}\label{V*M_nx}
\frac{1}{2}\mathcal{V}^{(0,n)} \mathcal{M}^{(0,n)}x=x-\sum_{k=1}^{2m-1}(x,f_k)e_{n-k}
 +\mathcal{E}^{(0,n)T}x,\quad ||\mathcal{E}^{(0,n)T}x||\le e^{-C\log^2n}||x||,
\end{equation}

The idea of the proof is to show that for $|j-n|,|k-n|\le [n^{1/4}]$
\begin{equation}\label{dM}
M_{k,j-1}-M_{k,j+1}=M_{j+1,k}-M_{j-1,k}=2R_{k-j}+\e'_{j,k},\quad |\e'_{j,k}|\le
C_*n^{-1/9},
\end{equation}
where $R_k$ is defined by (\ref{R}).

If we know, e.g., $M_{n-1,n}$, these relations  allow us to find $M_{n+2j,n+1+2k}$,
going step by step from the point $(n-1,n)$ to $(n+2j+1,n+2k)$. Then, using the symmetry
  $M_{j,k}=-M_{k,j}$ we obtain $M_{n+2j,n+2k+1}$. Hence, since $M_{j,k}=0$
for even $j-k$  because of the evenness, we find in such a way all $M_{j,k}$ with
$|j-n|,|k-n|\le [n^{1/4}]$.
Thus, if we denote  $C(n)=M_{n-1,n}-M_2$ (see (\ref{M_k}) for the definition of $M_2$), then
for odd $j-k$ we have
\begin{equation}\label{M,C}
M_{j,k}= M_{j,k}^*+\e_{j,k},\quad
   M_{j,k}^*=M_{k-j+1}-\frac{1}{2}((1+(-1)^j)M_{-\infty}-(-1)^jC(n),
\end{equation}
.
\begin{equation}\label{M_k}\begin{array}{l}
\displaystyle   M_{k}=(1+(-1)^k)\sum_{j=k}^\infty R_{j}=P^{-1}(2)-\frac{1}{2\pi}
\int_{-\pi}^{\pi}   \frac{\sin(k-1)x\, dx}{P(2\cos x)\sin x }\\
\displaystyle M_{-\infty}= 2\sum_{j=-\infty}^\infty R_{j}=2P^{-1}(2),
\end{array}\end{equation}
where $P$ is defined in (\ref{P}), and
\begin{equation}\label{b_e}\begin{array}{l}
\displaystyle|\e_{2j-1,2k}|\le |\e_{n-1,n}'|+|\sum_{j'\in[j,n/2]}|\e_{2j'-1,2k}'|+
\sum_{k'\in[k,n/2]}|\e_{n-1,2k}'|\\
\le C_*n^{-1/9}(1+|j-n|+|k-n|).\end{array}\end{equation}
Remark, that the expression for $M_{j,k}$ in the case $V(\lambda)=\lambda^{2p}+o(1)$
were obtained in \cite{DeG1}.

Let us assume that we know (\ref{dM}) and show, how the assertion of Theorem
\ref{thm:1} can be obtained from (\ref{dM}).
The first step is
\begin{proposition}\label{p:loc}
Suppose   $\mathcal{M}^{(0,n)}x=i\varepsilon_0x$ ($||x||=1$).
Then there exists a vector $x_0\in \mathcal{I}^{(n-6m+2,n)}$ such that
\begin{equation}\label{pl.1}
    ||x_0-x||\le ||\mathcal{V}^{(0,n)}||\cdot|\varepsilon_0 |\le C_V|\varepsilon_0 | ,\quad
    ||\mathcal{M}^{(0,n)} x_0||\le 2|\varepsilon_0 |,
\end{equation}
where $C_V=\max_{\lambda\in[-2-d_1/2, 2+d_1/2]}|V'(\lambda)|$.
\end{proposition}
The proposition allows us  to replace the eigenvector $x$, which
in principle can have nonzero components even for $|k-n|\sim n$, by
the vector $x_0$ whose components are zero for $|k-n|>6m+2$. Then
we can replace $\mathcal{M}^{(0,n)}x_0$ by $\mathcal{M}^{(n-\widetilde N,n)}x_0$
with $\widetilde N=6m+2+2[\log^2n]$ and then, using (\ref{dM}) replace
$\mathcal{M}^{(n-\widetilde N,n)}x_0$ by $\mathcal{M}^{*(n-\widetilde N,n)}x_0$
(for more details see below).

The next step is to prove that Proposition \ref{p:loc} and (\ref{dM}) imply
the following representation of $x_0$
\begin{equation}\label{*.6}\begin{array}{l}
x_0=c_1r_{1}+y,\quad
r_1=(\mathcal{R}^{(n-\widetilde N,n)})^{-1}e_{n-1},\quad
||y||\le C_3\varepsilon_1, \\
c_1\in \mathbb{C},\quad \left|\,|c_1|\cdot||r_1||-1\right|\le C_4\varepsilon_1
\end{array}\end{equation}
with $C_3, C_4$ depending only on
$C_V$ from (\ref{pl.1}) and $\delta_1$, $\delta_2$ of (\ref{b_R}). Here and below
\begin{equation}\label{ti_N,e1}
\widetilde N=6m+2+2[\log^2n]\quad \varepsilon_1=\max\{|\epsilon_0|;2C_*n^{-1/9}\widetilde N^2\},
\end{equation}
with   $C_*$ defined in (\ref{dM}) and we  use that  (\ref{b_R}) yields
\begin{equation}\label{b_R.1}
\delta_1\mathcal{I}^{(n-\widetilde N,n)}\le\mathcal{R}^{(n-\widetilde N,n)}\le
\delta_2\mathcal{I}^{(n-\widetilde N,n)},
\end{equation}
thus $(\mathcal{R}^{(n-\widetilde N,n)})^{-1}$ exists and $||(\mathcal{R}^{(n-\widetilde
N,n)})^{-1}||\le\delta_1^{-1}$.

Assume that we have proved (\ref{*.6}).
Recall  that $\mathcal{M}^{(0,n)}$
is a skew symmetric matrix of the even dimension with real entries. Hence,
 if  $i\e_0$  is its eigenvalue,  $-i\e_0$
is  its eigenvalue too (if $\e_0=0$, then this eigenvalue has multiplicity at least 2). Thus,
there exists  an eigenvector $x^{(1)}$ such that
$\mathcal{M}^{(0,n)}x^{(1)}=-i\e_0 x^{(1)}$ and
\begin{equation}\label{*.7}
(x,x^{(1)})=0.
\end{equation}
 Then, using (\ref{*.6}), we can
 conclude that there exists $x^{(1)}_0$ such that
 \[||x^{(1)}_0-x^{(1)}||\le
C_V|\e_0|, \quad x^{(1)}_0=c_1^{(1)}r_1+y^{(1)}\]
with the same $r_1$ and some $y^{(1)}$ and  $c_1^{(1)}$, such that
\[||y^{(1)}||\le C_3\varepsilon_1,\quad
\left|\,|c_1^{(1)}|\cdot||r_1||-1\right|\le C_4\varepsilon_1,\]
where $\varepsilon_1$ is defined in (\ref{ti_N,e1}). Hence, it is easy to see that
\[|(x^{(1)}_0,x_0)|\ge |c_1|\cdot |c_1^{(1)}|\cdot||r_1||^2- C_5\varepsilon_1\ge
1-C_6\varepsilon_1\]
where $C_6$ depends only on $\delta_1,\delta_2$ and $||\mathcal{V}^{(0,n)}||$.
On the other hand, it follows from (\ref{*.7}) and (\ref{pl.1}) that
\[|(x^{(1)}_0,x_0)|\le 2C_V\e_1+C_V^2\e_1^2\]
The last two inequalities give us the contradiction, if $|\e_0|\le C_0$, where $C_0$
is some constant depending  only on $\delta_1,\delta_2$ and $C_V$. Hence, we conclude
that $\mathcal{M}^{(0,n)}$ have no eigenvalues in the interval $[-iC_0,iC_0]$.

 Thus, we have proved that (\ref{*.6}) yields the assertion of Theorem \ref{thm:1}.
 Let us derive (\ref{*.6}) from (\ref{dM}) and Proposition \ref{p:loc}.

 Since $\mathcal{M}^{(n-\widetilde N,n)}=
\mathcal{I}^{(n-\widetilde N,n)}\mathcal{M}^{(0,n)}\mathcal{I}^{(n-\widetilde N,n)}$ and
$x_0\in\mathcal{I}^{(n-\widetilde N,n)}\mathcal{H}$,  in view of Proposition \ref{p:loc}  we
have
\[||\mathcal{M}^{(n-\widetilde N,n)}x_0||\le||\mathcal{M}^{(0,n)}x_0|| \le 2|\varepsilon_0 |.\]
Hence, using (\ref{M,C}) and (\ref{b_e}) and (\ref{b_norm}), we have
\begin{multline*}
 ||\mathcal{M}^{*(n-\widetilde N,n)}x_0||\le||\mathcal{M}^{(n-\widetilde N,n)}x_0||
+||(\mathcal{M}^{(n-\widetilde N,n)}-\mathcal{M}^{*(n-\widetilde N,n)})x_0||\\
\le 2|\varepsilon_0| +2C_*n^{-1/9}\widetilde N^2
\le 3\varepsilon_1. \end{multline*}
Therefore, denoting $\widetilde y=\frac{1}{2}\mathcal{D}^{(n-\widetilde N,n)}\mathcal{M}^{*(n-\widetilde
N,n)}x_0$ and using that $||\mathcal{D}||\le 2$ (see (\ref{b_norm})), we have
\[||\widetilde y||=\frac{1}{2}||\mathcal{D}^{(n-\widetilde N,n)}\mathcal{M}^{*(n-\widetilde N,n)}x_0||
\le 3\varepsilon_1 \]
On the other hand, using the definition (\ref{M_k}), we get
\begin{equation}\label{DM}
\widetilde y=\frac{1}{2}\mathcal{D}^{(n-\widetilde N,n)}\mathcal{M}^{*(n-\widetilde N,n)}x_0=
\mathcal{R}^{(n-\widetilde N,n)}x_0-(x_0,\mu_1)e_{n-1}+(x_0,\mu_{\widetilde N})e_{n-\widetilde N},
\end{equation}
where  $\mu_1,\mu_{\widetilde N}\in\mathcal{I}^{(n-\widetilde N,n)}\mathcal{H}$ and have the components
$\mu_{1j}=M_{n,i}^*$, $\mu_{\widetilde Nj}=M_{n-\widetilde N-1,j}^*$, $i=n-\widetilde N,\dots,n-1$.
Note, that the second equality in (\ref{DM}) is valid for any $x$, not only for $x=x_0$.
If we apply $(\mathcal{R}^{(n-\widetilde N,n)})^{-1}$ to both sides of the above
equality and denote
\begin{equation}\label{c,r}\begin{array}{l}
 r_{\widetilde N}=(\mathcal{R}^{(n-\widetilde N,n)})^{-1}e_{n-\widetilde N},\quad
\widetilde y_0=(\mathcal{R}^{(n-\widetilde N,n)})^{-1}\widetilde y,\\
c_1=(x_0,\mu_1),\quad c_{\widetilde N}=-(x_0,\mu_{\widetilde N}) ,
 \quad \rho=||r_1||=||r_{\widetilde N}||,
\end{array}\end{equation}
 we get (cf (\ref{*.6}))
\begin{equation}\label{*.1}
x_0=c_1r_1+c_{\widetilde N}r_{\widetilde N}+\widetilde y_0,
 \quad
 ||\widetilde y_0||\le 3\varepsilon_1\cdot||\mathcal{R}^{(n-\widetilde N,n)})^{-1}||
 \le 3\varepsilon_1 \delta_1^{-1}
\end{equation}
and  by the definition (\ref{c,r}) and (\ref{b_R.1}) we have
$\delta_2^{-1}\le \rho\le \delta_1^{-1}$.

To proceed further we need some facts from the theory of Jacobi matrices.
\begin{proposition}\label{pro:*} Let  $J$ be a Jacobi matrix with
entries $|J_{j,j+1}|\le 1+d_1/4$ and $Q$ be a bounded analytic function
($|Q(z)|\le C_0$) in $\Omega[d_1/2,d_2/2]$. Then:

\noindent(i) for any $j,k$
\begin{equation}\label{p*.1}
|Q(\mathcal{J})_{j,k}|\le Ce^{-d|j-k|};
\end{equation}
\noindent(ii) if $\widetilde{\mathcal{J}}$ is another Jacobi matrix, satisfying the same
conditions, then
for any $j,k\in [n_1,n_2)$
\begin{multline}\label{p*.2}
|Q(\mathcal{J})_{j,k}-Q(\widetilde{\mathcal{J}})_{j,k}|\le \\C(\sup_{i\in[n_1,n_2)}|J_{i,i+1}
-\widetilde J_{i,i+1}|
e^{-d|j-k|}+e^{-d(|n_1-j|+|n_1-k|)}+e^{-d(|n_2-j|+|n_2-k|)});
\end{multline}
\noindent(iii) if $Q(\lambda)>\delta>0$ for
 $\lambda\in[-2-d_1/2,2+d_1/2]$, then for $i,j\in[n_1,n_2)$
\begin{multline}\label{p*.3}
|(Q(\mathcal{J})^{(n_1,n_2)})^{-1}_{j,k}-Q^{-1}(\mathcal{J})_{j,k}|\\
\le C\min\left\{e^{-d|n_1-j|}+e^{-d|n_2-j|},e^{-d|n_1-k|}+e^{-d|n_2-k|}\right\},\\
|(Q(\mathcal{J})^{(n_1,n_2)})^{-1}_{j,k}-(Q(\mathcal{J})^{(-\infty,n_2)})^{-1}_{j,k}|
\le C\min\left\{e^{-d|n_1-j|},e^{-d|n_1-k|}\right\},\end{multline}
where $C$ and $d$ depend only on $d_1,d_2,C_0$ and $\delta$.
\end{proposition}
The proof of Proposition \ref{pro:*} is given at the end of Section 3.

Using the exponential bounds (\ref{p*.3}) and (\ref{p*.1}), we obtain
 that the definition of $r_1$ and $r_{\widetilde N}$ (see (\ref{*.6}) and (\ref{c,r})) imply
\begin{equation}\label{*.3}
|(r_{1},r_{\widetilde N})|=\bigg|\sum_{j=n-\widetilde N}^{n-1}
(\mathcal{R}^{(n-\widetilde N,n)})^{-1}_{n-1,j}
(\mathcal{R}^{(n-\widetilde N,n)})^{-1}_{n-\widetilde N,j}
\bigg|\le C\widetilde N e^{-c\widetilde N}
\end{equation}
Hence, taking $(x_0,x_0)$, we obtain from (\ref{*.1})
\begin{multline*}
(|c_1|^2+|c_{\widetilde N}|^2)\rho^2-
(|c_1|^2+|c_{\widetilde N}|^2)^{1/2}\rho\left(||\widetilde y_0||+O(e^{-c\widetilde N/2})\right)-||\widetilde
y_0||^2\le
(x_0,x_0)\\
\le (|c_1|^2+|c_{\widetilde N}|^2)\rho^2+
(|c_1|^2+|c_{\widetilde N}|^2)^{1/2}\rho\left(||\widetilde y_0||+O(e^{-d\widetilde N/2})\right)+
||\widetilde y_0||^2.
\end{multline*}
Note that these inequalities are quadratic with respect to
$(|c_1|^2+|c_{\widetilde N}|^2)^{1/2}\rho$. Hence, using that  $(x,x)=1$ and
so, by (\ref{pl.1}), $|(x_0,x_0)-1|\le 2C_V|\varepsilon_1+C_V^2\varepsilon_1^2$, we obtain
  that there exists $C_1$ depending only on $\delta_1,
\delta_2,C_V$, such that
\begin{equation}\label{*.4}
1-C_1\varepsilon_1 \le (c_1^2+c_{\widetilde N}^2)^{1/2}\rho\le 1+C_1\varepsilon_1.
\end{equation}
On the other hand, since by Proposition \ref{p:loc}
$x_0\in \mathcal{I}^{(n-6m+2,n)}\mathcal{H}$, similarly to (\ref{*.3}),
we have from the exponential bounds (\ref{p*.3}) and (\ref{p*.1}) that
\[(x_0, r_{\widetilde N})=O(e^{-c\log^2n}).\]
Thus,  taking the scalar product of the r.h.s. of (\ref{*.1}) with
$r_{\widetilde N}$,
 we get   from (\ref{*.3})  that there exist $C_2$ depending only on $\delta_1,
\delta_2,C_V$, such that
\[
|c_{\widetilde N}|\le C_2\varepsilon_1.
\]
Collecting the above bounds, we obtain  (\ref{*.6}).

To finish the proof of Theorem \ref{thm:1} we are left to prove (\ref{dM}).
Define $\mathcal{V}^*=\{\mathcal{V}_{j,l}^*\}_{j,l=-\infty}^\infty$
with $\mathcal{V}^*_{j,l}=\hbox{sign}(l-j)V'(\mathcal{J}^*)_{j,l}$, where
$\mathcal{J}^*$ is defined in (\ref{J^*}). Then by the spectral theorem
\begin{equation}\label{nu*}\mathcal{V}^*_{j,l}=\mathcal{V}^*_{j-l}
=
\fracd{\hbox{sign}(l-j)}{2\pi}\intd_{-\pi}^{\pi} dx V'(2\cos x)e^{i(j-l)x}.
\end{equation}

The key point in the proof  of (\ref{dM}) is the lemma:
\begin{lemma}\label{lem:2} Under conditions of  Theorem \ref{thm:1}
\begin{equation}\label{fact}
  \mathcal{V}^*=\mathcal{P}\mathcal{D}=\mathcal{D}\mathcal{P},
\end{equation}
where $\mathcal{P}$ is defined in (\ref{calP}) and
\begin{equation}\label{d^*}
\mathcal{D}=\{D_{j,k}\}_{j,k=-\infty}^{\infty},\quad D_{j,k}=\delta_{j+1,k}-\delta_{j-1,k}.
\end{equation}
Moreover, for $N=2[n^{1/4}]$ there exist some error matrices $\widetilde{\mathcal{P}}$ and
$\widetilde{\mathcal{D}}$,
whose entries are equal to zero if $|j-n|>N+2m$, or $|k-n|>N+2m$, or $ |j-k|>2m-2$, admit the bounds
\begin{equation}\label{ti_P,D}\begin{array}{l}
|\ti P_{j,k}|\le CNmn^{-1},\\
\ti D_{j,k}=\delta_{j+1,k}d_{j+1},\quad
 |d_j|\le Cm^2n^{-1},
\end{array}\end{equation}
and   satisfy the relation
\begin{equation}\label{apr_fact}
  \mathcal{V}_{j,k}=(\mathcal{D}+\widetilde{\mathcal{D}})(\mathcal{P}+\widetilde{\mathcal{P}})_{j,k}
  +\widetilde\varepsilon_{j,k},\quad
 |k-n|\le N,\,\, | j-n|\le N+2m,
  \end{equation}
where $\widetilde\varepsilon_{j,k}=0$, if $|j-k|>2m-1$ and
\begin{equation}\label{var_j,k}
|\widetilde\varepsilon_{j,k}|\le CNm^4n^{-2},\quad if \quad |j-k|\le 2m-1.
\end{equation}
\end{lemma}
\begin{remark} Let us note  that if we write (\ref{fact}) for components and use
(\ref{lim_V}) and (\ref{p*.2}), we obtain for $n-N+1\le j,k\le n+N-1,$
\[\mathcal{V}_{j,k}=(\mathcal{P}\mathcal{D})_{j,k}+O(Nmn^{-1}).\]
It would be possible to use  this representation of $\mathcal{V}_{j,k}$ instead of
(\ref{apr_fact}),  if we know from the very beginning
that $||\epsilon\psi^{(n)}_k||_2=O(n^{-1/2})$. The last relation in principle could be obtained
 from the results of \cite{De:98} on the asymptotic of orthogonal polynomial
$p^{(n)}_k$ for $n-N+1\le k\le n+N-1$. But from our point of view it is more simple
to prove   more precise relations  (\ref{apr_fact})-(\ref{var_j,k}), which make it possible to
prove (\ref{dM}) without integration of the asymptotic of \cite{De:98}. One more reason
do not use \cite{De:98} is to make our proof applicable to non analytic potentials $V$, for
which \cite{De:98} does not work.\end{remark}

Since $\e$ is a bounded operator in $L_2[-2-d/2,2+d/2]$ by (\ref{nu*psi}),
(\ref{apr_fact}), (\ref{var_j,k}) and (\ref{nu*psi}) we have for $|k-n|< N$
\begin{equation}\label{t2.1}
\frac{1}{2}\sum_{j}(\mathcal{P}^{(0,\infty)}+\widetilde{\mathcal{P}})_{j,k}\left((1+d_{j})\epsilon\psi^{(n)}_{j-1}
-\epsilon\psi^{(n)}_{j+1}\right)=n^{-1}\psi^{(n)}_k+r_k,
\end{equation}
where $O_2(.)$ is defined in (\ref{psi'}) and for $|k-n|< N$
\begin{equation}\label{t2.2}
r_k:=\frac{1}{2}\sum_{|j-k|\le 2m-1}\widetilde\varepsilon_{j,k}
\epsilon\psi^{(n)}_j+O_2(e^{-C\log^2 n})=
O_2(Nm^5n^{-2}).
\end{equation}
Let us extend (\ref{t2.1}) to all $0\le k<\infty$, choosing  $r_k$ for $|k-n|\ge N$
in such a way to obtain for these $k$ identical equalities:
\begin{equation}\label{t2.2a}
r_k:=
\sumd_{j>0}(\mathcal{P}^{(0,\infty)}+\widetilde{\mathcal{P}})_{j,k}
\left((1+d_{j})\epsilon\psi^{(n)}_{j-1}
-\epsilon\psi^{(n)}_{j+1}\right) -\frac{1}{n}\psi^{(n)}_k =O_2(1),
\end{equation}
 Since $\mathcal{P}>\delta_2^{-1}$(see (\ref{b_R}))
 $\mathcal{P}^{(0,\infty)}>\delta_2^{-1}\mathcal{I}^{(0,\infty)}$.
 Moreover, by \ref{b_norm}) and (\ref{ti_P,D}),
 \[||\widetilde{\mathcal {P}}||\le
 CNm^2n^{-1}. \]
 Hence,  $\mathcal{P}^{(0,\infty)}+\widetilde{\mathcal{P}}$ has an
 inverse operator bounded uniformly in $n$.
Then, using twice the resolvent identity
\begin{equation}\label{res_id}
H_1^{-1}-H_2^{-1}=H_1^{-1}(H_2-H_1)H_2^{-1}
\end{equation}
for $H_1=\mathcal{P}^{(0,\infty)}$,
$H_2=\mathcal{P}^{(0,\infty)}+\widetilde{\mathcal{P}}$,
we obtain that
\begin{multline*} (\mathcal{P}^{(0,\infty)}+\widetilde{\mathcal{P}})^{-1}=
(\mathcal{P}^{(0,\infty)})^{-1}-
(\mathcal{P}^{(0,\infty)})^{-1}\widetilde{\mathcal{P}}(\mathcal{P}^{(0,\infty)})^{-1}\\+
(\mathcal{P}^{(0,\infty)})^{-1}\widetilde{\mathcal{P}}(\mathcal{P}^{(0,\infty)})^{-1}
\widetilde{\mathcal{P}}(\mathcal{P}^{(0,\infty)}+\widetilde{\mathcal{P}})^{-1}.
\end{multline*}
Hence,
\begin{multline}\label{t2.3}
\frac{1}{2}\left((1+d_{j})\epsilon\psi^{(n)}_{j-1}
-\epsilon\psi^{(n)}_{j+1}\right)=\sum_{k>0}
\left((\mathcal{P}^{(0,\infty)}+\widetilde{\mathcal{P}})^{-1}\right)_{k,j}(n^{-1}\psi^{(n)}_k+r_k)
\\
= n^{-1}\sum_{k>0}(\mathcal{P}^{(0,\infty)})^{-1}_{k,j}\psi^{(n)}_k+\sum_{k>0}
 \left((\mathcal{P}^{(0,\infty)})^{-1}(I-\widetilde{\mathcal{P}}
 (\mathcal{P}^{(0,\infty)})^{-1})\right)_{k,j}r_k\\+O_2(n^{-1}||\widetilde{\mathcal{P}}||)
 +O_2(||\widetilde{\mathcal{P}}||^2)=n^{-1}\Sigma_{1j}+\Sigma_{2j}+O_2(||\widetilde{\mathcal{P}}||^2).
\end{multline}
Using (\ref{p*.3}) and (\ref{t2.2}), it is easy to obtain that uniformly
in  $|j-n|\le N/2$
\[||\Sigma_{2j}||_2\le C\sup_{|k-j|\le N/2}||r_k||_2+
Ce^{-cN}\sup_{k}||r_k||_2\le CNm^4n^{-2}.\]
Besides, it follows from (\ref{p*.3}) that
 uniformly in $|j-n|\le N/2$
 \[\Sigma_{1j}-\sum_{k>0}\mathcal{P}^{-1}_{j,k}\psi^{(n)}_k(\lambda)=O_2(e^{-cn}).\]
 Hence, we get from (\ref{t2.3}) that
\begin{equation}\label{dif_eps}
(1+d_{j})\epsilon\psi^{(n)}_{j-1}
-\epsilon\psi^{(n)}_{j+1}=2n^{-1}\sum_{k>0}{R}_{j,k}\psi^{(n)}_k
+O_2(N^2m^4n^{-2}).
\end{equation}
where $d_j$ is defined in (\ref{ti_P,D}) and  ${R}_{j,k}$ is defined in (\ref{R}).
\begin{proposition}\label{pro:1}
There exists  $j:n\le j<n+N/2$ ($N=[n^{1/4}]$) such that
\begin{equation}\label{Dir}
||\epsilon\psi^{(n)}_{j}||^2_2+||\epsilon\psi^{(n)}_{j-1}||^2_2\le CN^{-1}.
\end{equation}
\end{proposition}
Using the proposition and (\ref{dif_eps}), we obtain by induction that
(\ref{Dir}) holds with some $C_1$ for all $j:|j-n|\le N/2$. Hence (\ref{dif_eps})
yields for all $j:|j-n|\le N/2$
\[
\epsilon\psi^{(n)}_{j-1}
-\epsilon\psi^{(n)}_{j+1}=2n^{-1}\sum_{k>0}{R}_{j,k}\psi^{(n)}_k
+O_2(m^2n^{-1}N^{-1/2}),
\]
where we used the bound (\ref{ti_P,D}) for $d_j$ and also that $N^2m^4n^{-2}\le
m^2n^{-1}N^{-1/2}$.
Multiplying the relation by $\psi^{(n)}_{k}$ and using (\ref{M}), we get (\ref{dM}).

The assertion of Theorem \ref{thm:1} is proved. In addition let us prove that
the constant $C(n)$ from (\ref{M_k}) tends to zero, as $n\to\infty$. To this end
consider the matrix $\mathcal{M}^{(0,n+1)}$. It is a skew symmetric matrix of the
odd dimension, hence it has at least one zero eigenvalue. Let $x$ be a corresponding
eigenvector. Applying $\mathcal{V}^{(0,n+1)}\mathcal{M}^{(0,n+1)}x=0$ we conclude that
$x\in\mathcal{I}^{(n-2m+2,n+1)}\mathcal{H}$. Then repeating the arguments, using in the
proof of (\ref{*.6}), we obtain that
\[\begin{array}{l}
||\mathcal{D}^{(n-\widetilde N,n+1)}\mathcal{M}^{(n-\widetilde N,n+1)}x||\le \e_1, \\
 x=c_1(\mathcal{R}^{(n-\widetilde N,n+1)})^{-1}e_n+y,\quad ||y||\le \e_1.\end{array}\]
Hence, using (\ref{DM}) for $x$, represented as above, we get
\[\begin{array}{l}
\dfrac{1}{2}\mathcal{D}^{(n-\widetilde N,n+1)}\mathcal{M}^{(n-\widetilde N,n+1)}x=
\left(1-((\mathcal{R}^{(n-\widetilde N,n+1)})^{-1}e_n,\mu_0)\right)c_1e_n+\tilde y\\
  ||\tilde y||\le C\e_1,
\end{array}\]
 where $\widetilde N$, $\varepsilon_1$ are defined in (\ref{ti_N,e1}),
  $\mu_0\in\mathcal{I}^{(n-\widetilde N,n+1)}\mathcal{H}$ and
its components for $i=0,2,\dots,n-\widetilde N$ are $\mu_{0,i}=M^*_{n+1,n-i}=M_{-i}+C(n)$
(see (\ref{M_k})). Hence, we conclude that
\begin{equation}\label{*.8}
\Delta_n:=|1-((\mathcal{R}^{(n-\widetilde N,n+1)})^{-1}e_n,\mu_0)|\le C\e_1
\end{equation}
But it follows from (\ref{p*.1}) and (\ref{p*.3}), that we can consider only the last
(2m-1) components of $\mu_0$
\[|((\mathcal{R}^{(n-\widetilde N,n+1)})^{-1}e_n,\mu_0)-
((\mathcal{R}^{(n-\widetilde N,n+1)})^{-1}e_n,\mathcal{I}^{(n-2m,n+1)}\mu_0)|\le
e^{-cm}=e^{-c\log^2m}.\]
On the other hand, the definition of $\mu_{0,i}$ and (\ref{M,C}), combined with
 (\ref{p*.1}) and (\ref{p*.3}) yield
\[
\mathcal{I}^{(n-2m,n+1)}\mu_0=C(n)\sum_{i=0}^{m}e_{n-2i}+
\sum_{i=0}^{m}\mathcal{R}^{(n-\widetilde N,n+1)}e_{n-2i}+O(e^{-c\log^2n}).
\]
Using this representation in   (\ref{*.8}), we obtain  that
\begin{equation}\label{*.9}
\Delta_n=|C(n)\sum_{i=0}^{m}(\mathcal{R}^{(n-\widetilde N,n+1)})^{-1}_{n,n-2i}
+O(e^{-c\log^2n})|\le C\e_1
\end{equation}
\begin{lemma}\label{lem:3}
\begin{equation}\label{l3.1}
   \sum_{i=0}^{m}(\mathcal{R}^{(n-\widetilde N,n+1)})^{-1}_{n,n-2i}=
   \left[(\mathcal{R}^{(-\infty,n+1)})^{-1}_{n,n}\right]^{1/2}P^{1/2}(2)+O(e^{-c\log^2n}),
\end{equation}
\end{lemma}
Lemma \ref{lem:3} and (\ref{*.9}) give us
\[|C(n)|\le C\e_1\le C'n^{-1/9}\log^2n,\]
where $C'$ depends only on $C_V$ of  $\delta_1$, $\delta_2$ of (\ref{b_R}) and
$C_*$ of (\ref{M_k}).

$\square$

\begin{corollary}\label{c:1}
Under conditions of Theorem \ref{thm:1} for $n> j,k>n-2[\log^2 n]$
\begin{equation}\label{M^-1}
  (\mathcal{M}^{(0,n)})^{-1}_{j,k}=
  \frac{1}{2}((\mathcal{R}^{(-\infty,n)})^{-1}\mathcal{D}^{(-\infty,n)})_{j,k}
  +\frac{1}{2}b_ja_k+O(n^{-1/10}),\end{equation}
where
\begin{equation}\label{a,b}
a_k=((\mathcal{R}^{(-\infty,n)})^{-1}e_{n-1})_{k},\quad
b_j=((\mathcal{R}^{(-\infty,n)})^{-1}r^*)_{j},\quad
r^*_{n-i}=R_i
\end{equation}
 with $R_i$ defined by (\ref{R}).
  For other $j,k$
\begin{equation}\label{M^-1.1}
(\mathcal{M}^{(0,n)})^{-1}_{j,k}=\frac{1}{2}\mathcal{V}^{(0,n)}_{j,k}+O(e^{-c\log^2n})
\end{equation}
\end{corollary}
\noindent{\it Proof of Corollary \ref{c:1}.} According to
Theorem \ref{thm:1}, $||(\mathcal{M}^{(0,n)})^{-1}||\le C$, with some
$n$-independent $C$, hence, applying $(\mathcal{M}^{(0,n)})^{-1}$ from the left
to both sides of (\ref{M_n*V}) and using that $(\mathcal{M}^{(0,n)})^{-1}$ is a skew
symmetric matrix, we obtain  (\ref{M^-1.1}) and prove that
$(\mathcal{M}^{(0,n)})^{-1}$ has not more than $(4m-1)$ nonzero diagonals. Thus,
it follows from the standard linear algebra arguments, that if we consider
$\mathcal{M}^{(n-\widetilde N,n)}$ with $\widetilde N$ defined in (\ref{ti_N,e1}),
then  $\widetilde N-2m$  last columns of $(\mathcal{M}^{(0,n)})^{-1}$
coincide with that of $(\mathcal{M}^{(n-\widetilde N,n)})^{-1}$, if it exists.
But according to (\ref{b_e})  and (\ref{b_norm}), if there exists
$(\mathcal{M}^{*(n-\widetilde N,n)})^{-1}$, and
$||(\mathcal{M}^{*(n-\widetilde N,n)})^{-1}||\le C$ with some $C$ independent of $n$,
then $(\mathcal{M}^{(n-\widetilde N,n)})^{-1}$ also exists and
\[||(\mathcal{M}^{(n-\widetilde N,n)})^{-1}-\mathcal{M}^{*(n-\widetilde N,n)})^{-1}||
\le C'n^{-1/9}\log^2n,\]
Thus, for our goal it is enough to study $(\mathcal{M}^{*(n-\widetilde N,n)})^{-1}$.
Using (\ref{DM}), we have
\begin{equation}\label{c1.1}
\mathcal{D}^{(n-\widetilde N,n)}\mathcal{M}^{*(n-\widetilde N,n)}=
\mathcal{R}^{(n-\widetilde N,n)}\left(\mathcal{I}^{(n-\widetilde N,n)}-\Pi^{(n-\widetilde
N,n)}_1+\Pi^{(n-\widetilde N,n)}_{\widetilde N}\right),
\end{equation}
where $\Pi^{(n-\widetilde N,n)}_1$ and $\Pi^{(n-\widetilde N,n)}_{\widetilde N}$ are
rank one matrices of the form
\[\begin{array}{l}
\Pi^{(n-\widetilde N,n)}_1x=(\mu_1,x)(\mathcal{R}^{(n-\widetilde N,n)})^{-1}e_{n-1},\\
\Pi^{(n-\widetilde N,n)}_{\widetilde N}x=
(\mu_{\widetilde N},x)(\mathcal{R}^{(n-\widetilde N,n)})^{-1}e_{n-\widetilde N},
\end{array}\]
where $\mu_1$ and $\mu_n$ are defined in (\ref{DM}).
Since $(\mathcal{R}^{(n-\widetilde N,n)})^{-1}e_{n-\widetilde N}$ and $\mu_{\widetilde
N}$ have zero odd components, while $((\mathcal{R}^{(n-\widetilde N,n)})^{-1}e_{n-1}$
and $\mu_1$ -- have zero even ones,
\[(\mu_1,(\mathcal{R}^{(n-\widetilde N,n)})^{-1}e_{n-\widetilde N})=
(\mu_{\widetilde N},(\mathcal{R}^{(n-\widetilde N,n)})^{-1}e_{n-1})=0.\]
Thus, using the standard linear algebra arguments, we obtain that
the matrix $(\mathcal{I}^{(n-\widetilde N,n)}-\Pi^{(n-\widetilde
N,n)}_1+\Pi^{(n-\widetilde N,n)}_{\widetilde N})$ has the inverse matrix bounded
uniformly in $n$ iff
\begin{equation}\label{c1.2} |\Delta_n|:=|1-
(\mu_{1},(\mathcal{R}^{(n-\widetilde N,n)})^{-1}e_{n-1})|=|1+
(\mu_{\widetilde N},(\mathcal{R}^{(n-\widetilde N,n)})^{-1}e_{n-\widetilde N})|
\ge C\end{equation}
Here the  last equality follows from (\ref{M_k}) and the symmetry of
$(\mathcal{R}^{(n-\widetilde N,n)})^{-1}$. But using the same arguments, as in
the proof of $C(n)\to 0$, we obtain (cf (\ref{*.9}))
\[\Delta_n=
(M_{-\infty}+C(n))\sum_{i=0}^{m}(\mathcal{R}^{(n-\widetilde N,n)})^{-1}_{n-1,n-2i+1}
+O(e^{-c\log^2n})\]
Now, using that
\[(\mathcal{R}^{(n-\widetilde N,n)})^{-1}_{n-1,n-2i+1}=
(\mathcal{R}^{(n-\widetilde N+1,n+1)})^{-1}_{n,n-2i}=
(\mathcal{R}^{(n-\widetilde N,n+1)})^{-1}_{n,n-2i}+O(e^{-c\log^2n})\]
(see (\ref{p*.3}) for the last relation), by (\ref{l3.1}) we obtain (\ref{c1.2}) and
therefore it follows from (\ref{c1.1})
\begin{multline*}(\mathcal{M}^{*(n-\widetilde N,n)})^{-1}=
(\mathcal{R}^{(n-\widetilde N,n)})^{-1}\mathcal{D}^{(n-\widetilde N,n)}\\+
\Delta_n^{-1}(\Pi^{(n-\widetilde
N,n)}_1-\Pi^{(n-\widetilde N,n)}_{\widetilde N})
(\mathcal{R}^{(n-\widetilde N,n)})^{-1}\mathcal{D}^{(n-\widetilde N,n)}
\end{multline*}
Taking into account the definition of $\Pi^{(n-\widetilde
N,n)}_1$ and $\Pi^{(n-\widetilde N,n)}_{\widetilde N}$ we have
\begin{multline*}(\mathcal{M}^{*(n-\widetilde N,n)})^{-1}x=
((\mathcal{R}^{(n-\widetilde N,n)})^{-1}\mathcal{D}^{(n-\widetilde N,n)})x\\+
((\mathcal{R}^{(n-\widetilde N,n)})^{-1}e_{n-1},x)\nu+
((\mathcal{R}^{(n-\widetilde N,n)})^{-1}e_{n-\widetilde N},x)\nu',
\end{multline*}
where $\nu\in \mathcal{I}^{(n-2m,n)}\mathcal{H}$, $\nu'\in
\mathcal{I}^{(n-\tilde N,n-\tilde N+2m)}\mathcal{H}$ and $||\nu||,||\nu'||\le C
\widetilde N$. Then (\ref{p*.3}),  (\ref{p*.1}) and  imply that for
any $x\in\mathcal{I}^{(n-2m,n)}\mathcal{H}$ the last term of the above relation is
$O(e^{-c\log^2n})$. Hence, for any $x\in\mathcal{I}^{(n-2m,n)}\mathcal{H}$, $||x||\le 1$
\begin{multline}\label{c1.3}
 (\mathcal{M}^{*(n-\widetilde N,n)})^{-1}x=
 ((\mathcal{R}^{(-\infty,n)})^{-1}\mathcal{D}^{(-\infty,n)})x+
(a,x)\nu+O(e^{-c\log^2n}),
\end{multline}
where the vector $a$ is defined in (\ref{a,b})
and we use the notation $x_1=O(e^{-c\log^2n})$ for vector
$x_1\in\mathcal{I}^{(n-\widetilde N,n)}\mathcal{H}$, if all its components
 $x_{1i}=O(e^{-c\log^2n})$.

Making the transposition of both sides of the last equation (recall that
$\mathcal{M}^{*(n-\widetilde N,n)T}=-\mathcal{M}^{*(n-\widetilde N,n)}$
and $\mathcal{D}^{(-\infty,n)T}=-\mathcal{D}^{(-\infty,n)}$), we get
for any $x\in\mathcal{I}^{(n-2m,n)}\mathcal{H}$
\begin{multline}\label{c1.4}
 (\mathcal{M}^{*(n-\widetilde N,n)})^{-1}x=
 (\mathcal{D}^{(-\infty,n)}(\mathcal{R}^{(-\infty,n)})^{-1})x
-(x,\nu)a+O(e^{-c\log^2n})
\end{multline}
Subtracting (\ref{c1.3}) from (\ref{c1.4}) we have
\begin{equation}\label{comm}
 [(\mathcal{R}^{(-\infty,n)})^{-1},\mathcal{D}^{(-\infty,n)}]x=
(a,x)\nu+(\nu,x)a+O(e^{-c\log^2n})
\end{equation}
where the symbol $[.,.]$ means the commutator.

On the other hand, it is easy to see that
\[[\mathcal{D}^{(-\infty,n)},\mathcal{R}^{(-\infty,n)}]x=
(x,r^*)e_{n-1}+(x,e_{n-1})r^*,\]
where $r^*$ is defined in Corollary \ref{c:1}. Hence,
\[[(\mathcal{R}^{(-\infty,n)})^{-1},\mathcal{D}^{(-\infty,n)}]x
=(x,a)b+(x,b)a,
\]
with $a,b$ defined in  (\ref{a,b}). Using the last relation and (\ref{comm}), we
 obtain that for any $x\in\mathcal{I}^{(n-2m,n)}\mathcal{H}$
\begin{equation}\label{comm1}
(x,a)b+(x,b)a=(a,x)\nu+(\nu,x)a+O(e^{-c\log^2n}).
\end{equation}
 Taking an arbitrary $x$ such that $(a,x)=(b,x)=0$ we obtain that
 \[\nu=\lambda_1a+\lambda_2b+O(e^{-c\log^2n})\]
Substituting this expression in (\ref{comm1})
we obtain $\lambda_1=O(e^{-c\log^2n})$, $\lambda_2=1-O(e^{-c\log^2n})$.
This relations combined with (\ref{c1.3}) prove  (\ref{M^-1}).

$\square$

\noindent{\it Proof of Theorem \ref{thm:1}.}
Substituting (\ref{M^-1}) in (\ref{S}) and using (\ref{nu*psi}),
 we obtain
\begin{equation}\label{S.1}
S_n(\lambda,\mu)=K_{n,2}(\lambda,\mu)+nr_n(\lambda,\mu),
\end{equation}
where $K_{n,2}(\lambda,\mu)$ is defined by (\ref{k_n}) and
\begin{equation}\label{r_n}
r_n(\lambda,\mu)=\sum_{j,k=-2m+1}^{2m-1}r_{j,k}\psi^{(n)}_{n-j}(\lambda)
(\epsilon \psi^{(n)}_{n-j})(\mu), \quad |r_{j,k}|\le C.
\end{equation}
According to the result of \cite{Tr-Wi:98}, to prove the weak
convergence of all correlation functions it is enough to prove the
weak convergence of cluster functions, which have the form
\begin{equation}\label{cl}R_n(s_1,\dots, s_k)=
\frac{\hbox{Tr} K_{n,1}(\lambda_0+\frac{s_1}{n\rho(\lambda_0)},
\lambda_0+\frac{s_2}{n\rho(\lambda_0)})\dots K_{n,1}(\lambda_0+\frac{s_1}{n\rho(\lambda_0)},
\lambda_0+\frac{s_1}{n\rho(\lambda_0)})}{(n\rho(\lambda_0))^k},
\end{equation}
where the matrix kernel $ K_{n,1}(\lambda,\mu)$ has the form (\ref{K_n1}) with
\[Sd_n(\lambda,\mu)=-n^{-1}\frac{\partial}{\partial\mu}S_n(\lambda,\mu),\quad
IS_n(\lambda,\mu)=n\int\epsilon(\lambda-\lambda')S_n(\lambda',\mu)d\lambda'.\]
Define similarly
\begin{equation}\label{I,d}\begin{array}{l}
Kd_{n,2}(\lambda,\mu)=n^{-1}\fracd{\partial}{\partial\mu}K_{n,2}(\lambda,\mu),\quad
IK_{n,2}(\lambda,\mu)=n\int\epsilon(\lambda-\lambda')K_{n,2}(\lambda',\mu)d\lambda'\\
rd_{n}(\lambda,\mu)=-n^{-1}\fracd{\partial}{\partial\mu}r_{n}(\lambda,\mu),\quad
Ir_{n}(\lambda,\mu)=n\int\epsilon(\lambda-\lambda')r_{n}(\lambda',\mu)d\lambda'.
\end{array}\end{equation}
\begin{lemma}\label{lem:4} Under conditions of Theorem \ref{thm:1} uniformly in
$|k-n|\le n^{1/5}$
\begin{equation}\label{l4.0}
||\epsilon \psi^{(n)}_{k}||_2= O(n^{-1/2}),
\end{equation}
and for any $\delta>0$ there exist $C$
such that uniformly in $\lambda\in[-2+\delta,2-\delta]$
\begin{equation}\label{l4.1}
|\epsilon \psi^{(n)}_{k}(\lambda)|\le C\left[n^{-1}+(1-(-1)^k)n^{-1/2}\right].
\end{equation}
Moreover, for any compact $\mathbb{K}\subset\mathbb{R}$ uniformly in $s_1,s_2\in\mathbb{K}$
\begin{equation}\label{l4.2}
\bigg|\bigg(\frac{\partial}{\partial s_1}+\frac{\partial}{\partial s_2}\bigg)
K_{n,2}(\lambda_0+s_1/n,
\lambda_0+s_2/n)\bigg|\le C,
\end{equation}
\begin{equation}\label{l4.3}
n^{-1}IK_{n,2}(\lambda_0+s_1/(n\rho(\lambda_0)),
\lambda_0+s_2/(n\rho(\lambda_0)))\to\epsilon
K_{\infty,2}^{(0)}(s_1-s_2).
\end{equation}
\end{lemma}
Since  in (\ref{S.1})-(\ref{r_n}) $r_{j,k}=0$, if both  $j,k$ are odd,
the bounds (\ref{l4.1}) and relations (\ref{S.1})-(\ref{r_n}) yield that
 uniformly in $s_1,s_2\in\mathbb{K}$
\begin{equation}\label{t1.1}
\bigg|\bigg|\frac{1}{n}K_{n,1}(\lambda_0+\frac{s_1}{n\rho(\lambda_0)},
\lambda_0+\frac{s_2}{n\rho(\lambda_0)})-\frac{1}{n}\widetilde K_{n,1}
(\lambda_0+\frac{s_1}{n\rho(\lambda_0)},\lambda_0+\frac{s_2}{n\rho(\lambda_0)})\bigg|\bigg|
\le \frac{Cm^2}{n^{1/2}},
\end{equation}
where
\[\widetilde K_{n,1}(\lambda,\mu)=\left(\begin{array}{cc}K_{n,2}(\lambda,\mu)&
Kd_{n,2}S_n(\lambda,\mu)\\ IK_{n,2}(\lambda,\mu)-\epsilon(\lambda-\mu)
& K_{n,2}(\mu,\lambda)\end{array}\right)
\]
Hence, we can replace $ K_{n,1}$ by $\widetilde K_{n,1}$ in (\ref{cl}). Then,
using integration by parts and (\ref{l4.2}), we obtain that
\[I(a,b)=\int_a^b\dots\int_a^b R_n(s_1,\dots s_k)ds_1\dots ds_k\]
can be represented as a finite sum of the terms:
\begin{multline}
T(a,b;k_1,\dots,k_p;l_1,\dots,l_q)=\int_a^b\dots\int_a^bds_1\dots ds_k F_1(s_1,s_2)\dots
F_k(s_k,s_1)\\
\left(\delta(s_{k_1}-a)-\delta(s_{k_1}-b)+\dots+\delta(s_{k_p}-a)-\delta(s_{k_p}-b)\right),
\end{multline}
where
\[F_i(s,s')=\frac{1}{n\rho(\lambda_0)}\left\{\begin{array}{ll}IK_{n,2}(\lambda_0+\frac{s}{n\rho(\lambda_0)},
\lambda_0+\frac{s'}{n\rho(\lambda_0)})-\epsilon(s_1-s_2),&i=l_1.\dots,l_q,\\
K_{n,2}(\lambda_0+\frac{s}{n\rho(\lambda_0)},
\lambda_0+\frac{s'}{n\rho(\lambda_0)}),&
otherwise\end{array}\right.\] Using the result of \cite{PS:97} on
the universality for the unitary ensemble and (\ref{l4.3}) we can
take the limit $n\to\infty$ in each of these term.
Theorem \ref{thm:1} is proved.

\section{Auxiliary results}\label{sec:3}
{\it Proof of Proposition \ref{p:loc}}
According to (\ref{V*M_nx}) and the assumption of the proposition, we have
\begin{equation}\label{pl.2}\begin{array}{l}
\displaystyle
\mathcal{V}^{(0,n)}\mathcal{M}^{(0,n)} x=x-\sum_{i=1}^{2m-1}(f_i,x)e_{n-i}=
i\varepsilon_0 \mathcal{V}^{(0,n)}x\\
\displaystyle\Rightarrow
x=\sum_{i=1}^{2m-1}(f_i,x)e_{n-i}+i\varepsilon_0 \mathcal{V}^{(0,n)}x.
\end{array}\end{equation}
Denote
\begin{equation}\label{ti_x}
 \widetilde  x=i\varepsilon_0 \mathcal{V}^{(0,n)}\mathcal{I}^{(0,n-2m+1)}x  ,
   \quad  x_0=x-\widetilde x.
\end{equation}
Since $\mathcal{V}^{(0,n)}$ has only $4m-1$ nonzero diagonals,
 $\mathcal{V}^{(0,n)}\mathcal{I}^{(n-2m+1,n)}x\in\mathcal{I}^{(n-6m+2,n)}\mathcal{H}$.
Hence, using the second line of (\ref{pl.2}), we get
\[
x_0=\sum_{i=1}^{2m-1}(f_i,x)e_{n-i}+i\varepsilon_0 \mathcal{V}^{(0,n)}
\mathcal{I}^{(n-2m+1,n)}x\in\mathcal{I}^{(n-6m+2,n)}\mathcal{H}.
\]
Besides, evidently
\[||x-x_0||=||\widetilde  x||=
|\varepsilon_0 |\cdot||\mathcal{V}^{(0,n)}\mathcal{I}^{(0,n-2m+1)}x||\le
|\varepsilon_0 |C_V.\]
Moreover,
\[||\mathcal{M}^{(0,n)}\widetilde  x||=
|\varepsilon_0 |\cdot||\mathcal{M}^{(0,n)}\mathcal{V}^{(0,n)}\mathcal{I}^{(0,n-2m+1)}x||.\]
But, by (\ref{M_n*Vx}) for any $y\in\mathcal{I}^{(0,n)}\mathcal{H}$
\[\mathcal{M}^{(0,n)}\mathcal{V}^{(0,n)}y=y-\sum_{i=1}^{2m-1}(y,e_{n-i}) f_i\]
Applying this formula to $y=\mathcal{I}^{(0,n-2m+1)}x\in \mathcal{I}^{(0,n)}\mathcal{H}$ and taking into
account that $(\mathcal{I}^{(0,n-2m+1)}x,e_{n-i})=0$ for all $i=1,\dots, 2m-1$,
we have
\[||\mathcal{M}^{(0,n)}\widetilde  x||=|\varepsilon_0 |\cdot||\mathcal{I}^{(0,n-2m+1)}x||
\le|\varepsilon_0 | .\]
Hence,
\[||\mathcal{M}^{(0,n)}  x_0||=||\mathcal{M}^{(0,n)}(x-\widetilde
x)||\le||\mathcal{M}^{(0,n)}x||+||\mathcal{M}^{(0,n)}\widetilde x||\le 2|\varepsilon_0 |.\]
Proposition \ref{p:loc} is proved.

$\square$

{\it Proof of Lemma \ref{lem:2}.}
According to the standard theory of Toeplitz matrices
\[
\mathcal{V}^*_{k,j}=\mathcal{V}_{k-j}^*=\frac{1}{2\pi}\int_{-\pi}^{\pi}e^{i(k-j)x}
\widetilde{\mathcal{V}}(x)dx,
\]
where
\begin{equation}\label{ti-nu}
\widetilde{\mathcal{V}}(x)=2\sum_{k=1}^\infty V'_k\sin kx,\quad
V'_k=\frac{1}{2\pi}\int_{-\pi}^{\pi}e^{ikx}V'(2\cos x)dx,
\end{equation}
and to prove (\ref{fact}) it is enough to prove that
\begin{equation}\label{nu-P}
\widetilde{\mathcal{V}}(x)=2\sin x\cdot P(2\cos x).\end{equation}
Replacing in (\ref{P}) $z\to 2\cos x$, $2\cos
y\to(\zeta+\zeta^{-1})$, $dy\to (i\zeta)^{-1}d\zeta$ and using the Cauchy theorem, we get
\begin{multline*}
P(2\cos x)=
\frac{1}{2\pi i }\oint_{|\zeta|=1+\delta}\frac{V^{\prime
}(\zeta+\zeta^{-1} )-V^{\prime }(2\cos x)}{\zeta+\zeta^{-1}-2\cos x }\zeta^{-1}d\zeta =\\
\frac{1}{2\pi i }\oint_{|\zeta|=1+\delta}\frac{\sum V^{\prime}_k
(\zeta^k+\zeta^{-k} )d\zeta}{(\zeta-e^{ix})(\zeta-e^{-ix})}=
\frac{1}{2\pi i }\oint_{|\zeta|=1+\delta}\frac{\sum V^{\prime}_k
\zeta^kd\zeta}{(\zeta-e^{ix})(\zeta-e^{-ix})}\\=\sum_kV^{\prime}_k
\frac{\sin kx}{\sin x}=\frac{\widetilde{\mathcal{V}}(x)}{2\sin x}.
\end{multline*}
To prove (\ref{apr_fact}) it is enough to show that there exists $d_j$ and $\ti P_{j,k}$,
 satisfying (\ref{ti_P,D}) such that for $|j-n|\le N$
\begin{equation}\label{l2.1}
( \mathcal{ D}\widetilde{\mathcal{ P}}+\widetilde{\mathcal{ D}}\mathcal{ P})_{j,k}=(\mathcal{V}-
\mathcal{V}^*)_{j,k}.
\end{equation}
Then if we take $\widetilde\e_{j,k}=(\widetilde{\mathcal{ D}}\widetilde{\mathcal{
P}})_{j,k}$,   relations (\ref{apr_fact}) become identities and it follows from (\ref{ti_P,D})
that
\[|\widetilde\e_{j,k}|\le (4m-2)\max_{k-2m-1\le j \le k+2m-1}|d_j| |\widetilde P_{j+1,k}|
\le CNm^4n^{-2}.\]

Denote by $\Delta{\mathcal{V}}_{j,k}$ the r.h.s. of (\ref{l2.1}). Then
 for any fixed $k$ (\ref{l2.1})
is equivalent to the system of equations
\begin{equation}\label{l2.1a}\begin{array}{l}
\widetilde P_{j+1,k}-\widetilde P_{j-1,k}=\Delta{\mathcal{V}}_{j,k}-P_{k-j-1}d_{j+1},
\quad j=k-2m+1,\dots k+2m-1,\\
 \widetilde P_{k-2m,k}=\widetilde P_{k+2m,k}=0.
\end{array}\end{equation}
If we take the sum of the above relations for $j=k-2m+1,\dots l$, we get
\begin{equation}\label{l2.2}
\widetilde P_{l+1,k}=\sum_{j=k-2m+1}^l(\Delta{\mathcal{V}}_{j,k}-P_{k-j-1}d_{j+1}).
\end{equation}
It is evident that $\widetilde P_{l+1,k}$ for any $\{d_j\}$ satisfy (\ref{l2.1}),
may be except the condition $\widetilde P_{k+2m,k}=0$.
To satisfy this condition we need to have the equality
\begin{equation}\label{l2.2a}  \sum_{j=k-2m+1}^{k+2m-1}(\Delta{\mathcal{V}}_{j,k}-
P_{k-j-1}d_{j+1}) =
\sum_{j}(\Delta{\mathcal{V}}_{j,k}-
P_{k-j}d_{j})=0\end{equation}
Here we can take the sum over all $j$, since for $|j-k|>2m-1$,
 $\Delta{\mathcal{V}}_{j,k}=P_{k-j}=0$.  For $k=n-2N,\dots,n+2N$ and
 $j=n-N-2m,\dots,n+N+2m$ set
\begin{equation}\label{d_k}
  \widetilde v_k=\sum_{|j|\le 2m-1}\Delta{\mathcal{V}}_{k+j,k},\,\quad
  d_j=\sum_{|l-n|\le 2N}(\mathcal{P}^{(n-2N,n+2N)})^{-1}_{j,l}\widetilde v_l,.
\end{equation}
Then (\ref{l2.2a}) is evidently valid, so we find the solution of (\ref{l2.1a}).

To prove bounds (\ref{ti_P,D}) we note that since $\mathcal{V}^*_{k+j,k}=-\mathcal{V}^*_{k-j,k}$, we  have
\[|\widetilde v_k|\le 2m\max_{|j|\le 2m-1}|\mathcal{V}_{k+j,k}+\mathcal{V}_{k-j,k}|
= 2m\max_{|j|\le 2m-1}|\mathcal{V}_{k+j,k}-\mathcal{V}_{k,k-j}|\le Cm^2n^{-1}.
\]
Here the last bound follows from (\ref{p*.2}), if we take $\mathcal{J}=\mathcal{J}^{(n)}$,
define  $\widetilde{\mathcal{J}}$  by its coefficients $\widetilde{{J}}_{k,k+1}={J}^{(n)}_{k+j}$
and use (\ref{J_k}) to estimate  $|\widetilde{{J}}_{k,k+1}-{J}^{(n)}_{k}|$. Since
$(\mathcal{P}^{(n-2N,n+2N)})^{-1}$ is a bounded operator with coefficients satisfying
(\ref{p*.1})-(\ref{p*.3}),
from (\ref{d_k}) we obtain the bounds (\ref{ti_P,D}) for $\{d_j\}$.
Moreover, since  (\ref{J_k})  and (\ref{p*.2}) imply that
 $|\Delta{\mathcal{V}}_{j,k}|\le CNn^{-1}$, (\ref{l2.2}) yield
 bounds (\ref{ti_P,D}) for  $\widetilde P_{j,k}$.

$\square$

{\it Proof of Proposition \ref{pro:1}.}
Integrating by parts, we obtain that for any $k$
\begin{equation}\label{||ef||}
||\epsilon \psi^{(n)}_k||^2_2=-\frac{1}{4}\int_{-L}^L |\lambda-\mu|
\psi^{(n)}_k(\lambda)\psi^{(n)}_k(\mu)d\lambda d\mu+
\left(\mathbf{1}_{|\lambda|\le L},\psi^{(n)}_k\right)_2^2L/2
\end{equation}
with $L=2+d_1/2$. Therefore, using the Christoffel-Darboux formula, we get
\begin{multline*}
\sum_{n\le j\le n+N/2}(||\epsilon \psi_j^{(n)}||_2^2+||\epsilon \psi_{j-1}^{(n)}||_2^2)
\\
\le -2\int_{-L}^L\bigg(\left(\psi^{(n)}_{n+N/2+1}(\lambda)
\psi^{(n)}_{n+N/2}(\mu)-\psi^{(n)}_{n+N/2+1}(\mu)
\psi^{(n)}_{n+N/2}(\lambda)\right)J^{(n)}_{n+N/2+1}\\-\left(\psi^{(n)}_{n+1}(\lambda)
\psi^{(n)}_{n}(\mu)-\psi^{(n)}_{n+1}(\mu)
\psi^{(n)}_{n}(\lambda)\right)J^{(n)}_{n+1}\bigg)\hbox{sign}(\lambda-\mu)
 d\lambda d\mu\,+L||\mathbf{1}_{|\lambda|\le L}||^2_2\le C.
\end{multline*}
 Then, it is evident that there exists $n\le j\le n+ N/2$,
for which (\ref{Dir}) is valid.

$\square$

{\it Proof of Lemma \ref{lem:3}.}
Remark first that by (\ref{p*.3}) $(\mathcal{R}^{(n-\widetilde N,n+1)})^{-1}_{n,n-2i}$
in (\ref{l3.1}) can be replaced by $(\mathcal{R}^{(-\infty,n+1)})^{-1}_{n,n-2i}$ with the error
$O(e^{-c\log^2n})$.

Consider  ${ P}(\zeta+\zeta^{-1})\zeta^{2m-2}$ with $P$ defined in (\ref{P}).
It is easy to see that ${ P}(\zeta+\zeta^{-1})\zeta^{2m-2}$ is a polynomial of the
$(4m-4)$-th degree, which has the roots
$\{\zeta_j,\zeta_j^{-1}\}_{j=1}^{2m-2}$ with  $|\zeta_j|<1$ ($j=1,\dots,2m-2$). Denote
\begin{equation}\label{l3.3a}
 P_1(\zeta)
 =\prod_{j=1}^{2m-2}(\zeta-\zeta_j)=
 \sum_{j=0}^{2m-2}c_j\zeta^{2m-2-j} ,
  \quad
   P_2(\zeta)=\prod_{j=1}^{2m-2}(\zeta-\zeta_j^{-1}).
\end{equation}
Then
\begin{equation}\label{l3.3b}
P (\zeta+\zeta^{-1})=a_m\zeta^{-2m+2}P_1(\zeta)P_2(\zeta)
\end{equation}
with some positive $a_m$. Take $c_j$ from the representation
(\ref{l3.3a}). Using (\ref{R}) and (\ref{l3.3a}), we get
\begin{multline*}
\sum_{j=0}^{2m-2}R_{n-l,n-j}c_{j}= \frac{1}{2\pi
i}\oint\frac{\zeta^{-1}\sum_{j=0}^{2m-2}c_{j}\zeta^{l-j}d\zeta}{P(\zeta+\zeta^{-1})}\\
=\frac{1}{2\pi
i}\oint\frac{\zeta^{l-1}\sum_{j=0}^{2m-2}c_j\zeta^{2m-2-j}d\zeta}{a_mP_1(\zeta)P_2(\zeta)}
=\frac{1}{2\pi i}\oint\frac{\zeta^{l-1}d\zeta}{a_mP_2(\zeta)} =\frac{\delta_{0,l}}{a_mP_2(0)}.
\end{multline*}
Hence, we conclude that
\begin{equation}\label{l3.5}
 (R^{(-\infty,n+1)})^{-1}_{n,n-j}=a_mP_2(0)c_{j},\quad j=0,\dots
\end{equation}
In particular, we have
\[
\sum_{j=0}^{2m} (R^{(-\infty,n+1)})^{-1}_{n,n-j}=a_mP_2(0)P_1(1)
=(a_mP_2(0))^{1/2}(a_mP_1^2(1)P_2(0))^{1/2}.
\]
Using  the representation
\begin{equation}\label{P_i(1)}
 P(2)=a_mP_1(1)P_2(1)=a_m\prod_{j=1}^{2m-2}(1-\zeta_i)^2
 \prod_{j=1}^{2m-2}\zeta_i^{-1}
 =a_mP_1^2(1)P_2(0),
\end{equation}
and (\ref{l3.5}) for $j=1$ (recall that, according to (\ref{l3.3a}), $c_0=1$),
we obtain (\ref{l3.1}).

$\square$

{\it Proof of Lemma \ref{lem:4}.}
Using (\ref{||ef||}) and (\ref{rec}),  we get
\begin{multline}\label{l4.01}
||\epsilon\psi^{(n)}_k||^2_2=-\frac{1}{2}\int_{-L}^L\hbox{sign}(\lambda-\mu)(J^{(n)}_{k}
\psi^{(n)}_{k+1}(\lambda)+
J^{(n)}_{k-1}\psi^{(n)}_{k-1}(\lambda))\psi^{(n)}_k(\mu)d\lambda d\mu\\+
\frac{1}{2}(\psi^{(n)}_k,\mathbf{1}_{|\lambda|\le L})^2L
=-J^{(n)}_{k}M_{k+1,k}-
\J^{(n)}_{k-1}M_{k-1,k}+\frac{1}{2}(\psi^{(n)}_k,\mathbf{1}_{|\lambda|\le L})^2L.
\end{multline}
Hence due to (\ref{M,C})-(\ref{b_e}) we get (\ref{l4.0}) for odd $k$.
Besides, integrating by parts and using that  $\psi_{k}^{(n)}=(\epsilon\psi_{k}^{(n)})'$,
we obtain for odd $k$
\[\epsilon L\psi^{(n)}_k(\lambda)=\lambda\epsilon \psi^{(n)}_k(\lambda)
-\epsilon^2\psi^{(n)}_k(\lambda),
\]
where $L\psi_{k}(\lambda)=\lambda\psi_{k}(\lambda)$. Hence, using (\ref{rec}),
(\ref{l4.0}) and the bound $||\epsilon||\le (4+d)$, we get
\[||J^{(n)}_{k}\epsilon\psi^{(n)}_{k+1}+J^{(n)}_{k-1}\epsilon\psi^{(n)}_{k-1}||_2=
||\epsilon L\psi^{(n)}_k||_2\le C||\epsilon \psi^{(n)}_k||_2=O(n^{-1/2}).
\]
This relation, combined with (\ref{dif_eps}) prove (\ref{l4.0}) for even $k$.

To prove (\ref{l4.1}) we use the result of \cite{De:99a},
according to which, uniformly in any compact $\Delta\subset(-2,2)$
for $|k|\le 2m+1$
\begin{equation}\label{as_D}
\psi_{n+k}(\lambda)=\frac{2+\varepsilon_{n+k}}{\sqrt{2\pi}|4-\lambda^2|^{1/4}}
\cos\left(n\pi\int_\lambda^2\rho(\mu)d\mu+k\gamma(\lambda)+\frac{1}{2}\theta(\lambda)-
\frac{\pi}{4}\right)+O\left(m^2n^{-1}\right)
\end{equation}
where $\e_{n+k}\to 0$ does not depend on $\lambda$, $\rho(\lambda)$ is the limiting IDS,
 $\gamma(\lambda)$ is a smooth function in $(-2,2)$ and
$\cos\theta=\lambda/2$.

Integrating this  relations between $0$ and $\lambda$, we get
\[|\epsilon\psi_{n-k}^{(n)}(\lambda)-\epsilon\psi_{n-k}^{(n)}(0)|\le
Cm^2n^{-1}.\]
 Then, using the fact that
$(\epsilon f)(0)=0$ for
even $f$, we get (\ref{l4.1}) for even $k$
(recall, that $n$ is even). For odd $k$ the above inequality imply
\[||\epsilon\psi_{n-k}^{(n)}||_2\ge |\epsilon\psi_{n-k}^{(n)}(0)|+Cm^2n^{-1}.\]
Combining with (\ref{l4.0}), we get
(\ref{l4.1}).

Inequality (\ref{l4.2}) follows from the result \cite{PS:97} (see
Lemma 7), according to which
\begin{multline*}
\bigg|\bigg(\frac{\partial}{\partial s_1}+\frac{\partial}{\partial
s_2}\bigg) K_{n,2}(\lambda_0+s_1/n, \lambda_0+s_2/n)\bigg|\le C\bigg(n^{-1}|s_1-s_2|^2 \\
+ |\psi_{n}^{(n)}(\lambda_0+s_1/n)|^2+
|\psi_{n-1}^{(n)}(\lambda_0+s_1/n)|^2+
|\psi_{n}^{(n)}(\lambda_0+s_2/n)|^2+
|\psi_{n-1}^{(n)}(\lambda_0+s_2/n)|^2\bigg).\end{multline*}
 Since by(\ref{as_D}) $\psi_{n}^{(n)}$, $\psi_{n-1}^{(n)}$ are
 uniformly bounded in each compact $\mathbb{K}\subset (-2,2)$, we
 obtain (\ref{l4.3}).

To prove (\ref{l4.3}) we use  the Christoffel-Darboux formula, which gives us
\begin{multline}\label{l4.5}
n^{-1}IK_{n,2}(\lambda,\mu)=\int_{|\lambda'-\lambda_0|\ge\delta'}\epsilon(\lambda-\lambda')
\frac{\psi_{n}^{(n)}(\lambda')\psi_{n-1}^{(n)}(\mu)
-\psi_{n-1}^{(n)}(\lambda')\psi_{n}^{(n)}(\mu)}{\lambda'-\mu}d\lambda'\\
+\int_{|\lambda'-\lambda_0|\le\delta'}\epsilon(\lambda-\lambda')
\frac{\psi_{n}^{(n)}(\lambda')\psi_{n-1}^{(n)}(\mu)
-\psi_{n-1}^{(n)}(\lambda')\psi_{n}^{(n)}(\mu)}{\lambda'-\mu}d\lambda'=I_1+I_2
\end{multline}
Integrating by parts (we use again that $\psi_{k}^{(n)}=(\epsilon\psi_{k}^{(n)})'$)
 and taking into account that  $\epsilon\psi_{k}^{(n)}(\pm L)=
 \pm(\epsilon\psi_{k}^{(n)},\mathbf{1}_{|\lambda|\le L})=O(n^{-1/2})$ (see (\ref{l4.01})
 and (\ref{l4.0})), we get
\begin{multline*}
 |I_1|\le C\delta^{-1}n^{-1/2}
+\delta^{-2}\int_{-L}^L(|\epsilon\psi_{n}^{(n)}(\lambda')|+
|\epsilon\psi_{n-1}^{(n)}(\lambda')|)d\lambda'\\
\le C\delta^{-1}n^{-1/2}+C\delta^{-2}(||\epsilon\psi_{n-1}^{(n)}||_2+||\epsilon\psi_{n}^{(n)}||_2)=O(n^{-1/2}).
\end{multline*}
To find $I_2$ observe that (\ref{as_D}) yields for
$\lambda,\mu\in(-2+\e,2-\e)$
\begin{multline*}
n^{-1}K_{n,2}(\lambda,\mu)=R(\lambda)\frac{\sin\left( n\pi\int_{\mu}^{\lambda}\rho(\lambda')
d\lambda'\right)}{n(\lambda-\mu)}(1+(\lambda-\mu)\phi_1(\lambda,\mu))\\+ n^{-1}\cos\bigg(
n\pi\int_{\mu}^{\lambda}\rho(\lambda') d\lambda'\bigg)\phi_2(\lambda,\mu)+n^{-1}\cos\left(
n\pi(\int_{2}^{\lambda}+ \int_2^\mu)\rho(\lambda') d\lambda'\right)\phi_3(\lambda,\mu),
\end{multline*}
where $R$ and $\phi_1,\phi_2,\phi_3$ are  smooth functions of $\lambda$. Hence,
using  the Riemann-Lebesgue theorem to estimate integrals with
$\phi_i(\lambda,\mu)$, we obtain
\[I_2=\int_{-n\delta'}^{n\delta'}ds'\epsilon(s_1-s')R(\lambda_0+s'/n)
\frac{\sin\left(n\pi\int_{\lambda_0+s_2/n}^{\lambda_0+s'/n}\rho(\lambda')d\lambda'\right)}{s'-s_2}
+o(1).
\]
Now we split here the integration domain in two parts: $|s'|\le A$ and $|s'|\ge A$ and  take
 the limits
$n\to\infty$ and then $A\to\infty$.  Relation (\ref{l4.3}) follows.

$\square$

{\it Proof of Proposition \ref{pro:*}.}
Assertion (i) follows from the spectral theorem, according to which
\begin{equation}\label{p*.4}
Q(\mathcal{ J})_{j,k}= \frac{1}{2\pi
i}\oint_{d(z)=d}R_{j,k}(z)Q^{-1}(z)dz,
\end{equation}
and the bound, valid for the resolvent $R(z)=(\mathcal{ J}-z)^{-1}$
 of any Jacobi matrix $\mathcal{ J}$, satisfying conditions of the
proposition (see \cite{RS})
\begin{equation}\label{p*.5}
 |R_{a,b}|\le \frac{C}{d(z)}e^{-Cd(z)|a-b|},
\quad
d(z)=\hbox{dist}\,\{z,[-2-d_1/2,2+d_1/2]\}.
\end{equation}
To prove assertion (ii) consider
\[\mathcal{ J}(n_1,n_2)=\mathcal{ J}^{(n_1,n_2)}+\mathcal{ J}^{(-\infty,n_1)}
+\mathcal{ J}^{(n_2,\infty)},\quad \widetilde{\mathcal{ J}} (n_1,n_2)=\widetilde{
\mathcal{J}}^{(n_1,n_2)}+ \widetilde{\mathcal{ J}}^{(-\infty,n_1)} +\widetilde{\mathcal{
J}}^{(n_2,\infty)}\]
 and denote
\[
R^{(1)}(z)=(\mathcal{ J}(n_1,n_2)-z)^{-1},\quad
R^{(2)}(z)=(\widetilde{\mathcal{ J}} (n_1,n_2)-z)^{-1},\quad
\widetilde R(z)=(\widetilde{\mathcal{ J}}-z)^{-1}.
\]
It is evident that for $n_1\le j,k\le n_2$ and
$z\not\in[-2,2]$
\[
R^{(1)}_{j,k}(z)=(\mathcal{ J}^{(n_1,n_2)}-z)^{-1}_{j,k},\quad
R^{(2)}_{j,k}(z)=(\widetilde{\mathcal{ J}}^{(n_1,n_2)}-z)^{-1}_{j,k}.
\]
Then, using the resolvent identity (\ref{res_id}) and (\ref{p*.5}, we get
\[
|R^{(1)}_{j,k}(z)-R^{(2)}_{j,k}(z)|\le C\sup_{i\in[n_1,n_2)}|J_{i,i+1}
-\widetilde J_{i,i+1}|
\frac{e^{-d(z)|j-k|/2}}{d^2(z)}.
\]
On the other hand, by (\ref{res_id}) and (\ref{p*.5}), we obtain
\begin{multline*}
|R_{j,k}-R^{(1)}_{j,k}|\le|R_{j,n_1+1}R^{(1)}_{n_1,k}|+|R_{j,n_1}R^{(1)}_{n_1+1,k}|
+|R_{j,n_2}R^{(1)}_{n_2-1,k}|+|R_{j,n_2-1}R^{(1)}_{n_2,k}|\\
\le \frac{C}{d^2(z)}
(e^{-d(z)(|n_1-j|+|n_1-k|)}+e^{-d(z)(|n_2-j|+|n_2-k|)}).
\end{multline*}
Similar bound is valid for $|\widetilde R_{j,k}-R^{(2)}_{j,k}|$.
Then (\ref{p*.5}) and (\ref{p*.4}) yield (\ref{p*.2}).

To prove assertion (iii) observe that $x_j=(Q(\mathcal{J})^{(n_1,n_2)})^{-1}_{j,k}$ is the solution
of the infinite linear system:
\[\begin{array}{ll}
\sumd Q(\mathcal{J})_{i,j}x_j=\delta_{i,k}, & i\in[n_1,n_2)\\
\sumd Q(\mathcal{J})_{i,j}x_j=r_i:=
\sumd Q(\mathcal{J})_{i,j}(Q(\mathcal{J})^{(n_1,n_2)})^{-1}_{j,k}, & i\not\in[n_1,n_2).
\end{array}\]
Hence,
\[
(Q(\mathcal{J})^{(n_1,n_2)})^{-1}_{j,k}=Q^{-1}(\mathcal{J})_{j,k}
+\sum_{i\not\in[n_1,n_2)}Q^{-1}(\mathcal{J})_{j,i}r_i
\]
Now, using assertion (i), we obtain the first inequality in (\ref{p*.3}). For the second
the proof is the same.

\medskip
 {\bf Acknowledgements.} The author is grateful to Prof. L.Pastur for the
 fruitful discussion.
 The author  acknowledges also the INTAS Research Network 03-51-6637 for the
 financial support.


\small


 \begin{thebibliography}{99}

\bibitem{APS:97} Albeverio, S., Pastur, L., Shcherbina, M.: On Asymptotic
Properties of the Jacobi Matrix Coefficients.  Matem. Fizika,
Analiz, Geometriya {\bf 4}, 263-277 (1997)

\bibitem{APS:01} Albeverio, S., Pastur, L., Shcherbina, M.: On the $1/n$
expansion for some unitary invariant ensembles of random matrices.
 Commun. Math. Phys. {\bf 224},  271-305 (2001)

\bibitem{BPS} Boutet de Monvel, A., Pastur L., Shcherbina M.: On the
statistical mechanics approach in the random matrix theory.
Integrated density of states.  J. Stat. Phys. \textbf{79},
585-611 (1995)

\bibitem{CK} Claeys, T., Kuijalaars, A.B.J.: Universality of the double scaling limit
in random matrix models, preprint arXiv:math-ph/0501074

\bibitem{De:98}  Deift, P.,  Kriecherbauer,,T.,  McLaughlin,K. T.- R.: New
results on the equilibrium measure for logarithmic potentials in
the presence of an external field, J. Approx. Theory \textbf{95},
 388--475 (1998)


\bibitem{De:99} Deift, P., Kriecherbauer, T., McLaughlin, K., Venakides,
S., Zhou, X.: Uniform asymptotics for polynomials orthogonal with
respect to varying exponential weights and applications to
universality questions in random matrix theory. Commun. Pure
Appl. Math. \textbf{52}, 1335-1425 (1999)

\bibitem{De:99a} Deift, P., Kriecherbauer, T., McLaughlin, K., Venakides,
S., Zhou, X.: Strong asymptotics of orthogonal polynomials with
respect to exponential weights.  Commun. Pure Appl. Math.
\textbf{52}, 1491-1552 (1999)

\bibitem{DeG1} Deift, P., Gioev, D.: Universality in random matrix
theory for orthogonal and symplectic ensembles. Preprint arxiv:math-ph/0411075

\bibitem{DeG2} Deift, P., Gioev, D.: Universality at the edge of the spectrum
for unitary, orthogonal, and symplectic ensembles of random matrices
Preprint arxiv:math-ph/0507023

\bibitem{Dy} Dyson, D.J.: A Class of Matrix Ensembles.  J.Math.Phys.,\textbf{13}, 90-107 (1972)


\bibitem{Jo:98} Johansson, K.: On fluctuations of eigenvalues of random
Hermitian matrices. Duke Math. J. \textbf{91}, 151-204 (1998)



\bibitem{Me:91} M.L.Mehta, M.L.: \emph{Random Matrices}. New York: Academic
Press, 1991


\bibitem{PS:97} Pastur, L., Shcherbina, M.: Universality of the local
eigenvalue statistics for a class of unitary invariant random
matrix ensembles. \emph{J. Stat. Phys.} \textbf{86}, 109-147
(1997)
\bibitem{PS:03} Pastur, L., Shcherbina, M.:On the edge universality of the local
eigenvalue statistics of matrix models.  Matematicheskaya fizika,
analiz, geometriya \textbf{10}, N3, 335-365 (2003)

\bibitem{PS:07} Pastur, L., Shcherbina, M.:
Bulk universality and related properties of Hermitian matrix models
(in preparation)

\bibitem{RS}  Reed,M., Simon,B.:\emph{Methods of Modern Mathematical
Physics, Vol.IV}, Academic Press: New York, 1978

\bibitem{Sa-To:97}  Saff, E.,  Totik, V.: \emph{Logarithmic Potentials with
External Fields}. Springer-Verlag, Berlin, 1997

\bibitem{S:05} Shcherbina,M.: Double scaling limit
for matrix models with non analytic potentials. Preprint ArXiv:cond-mat/0511161

\bibitem{St1} Stojanovic, A.:  Universality in orthogonal and
symplectic invariant matrix models with quatric potentials.
Math.Phys.Anal.Geom. \textbf{3},  339-373 (2002)

\bibitem{St2} Stojanovic, A.:  Universalit\'{e} pour des
mod\'{e}les orthogonale ou symplectiqua et a  potentiel quartic.
Math.Phys.Anal.Geom. Preprint Bibos 02-07-98

\bibitem{Tr-Wi:98} C.A. Tracy, H. Widom: Correlation
functions, cluster functions, and spacing distributions for random
matrices. J.Stat.Phys. \textbf{92}, 809-835 (1998)

\end{thebibliography}
\end{document}